\documentclass[aps,pra,groupedaddress,twocolumn,10pt]{revtex4-1}

\usepackage{float}
\usepackage{graphicx}  
\usepackage{xcolor}
\usepackage{dcolumn}   
\usepackage{bm}        
\usepackage{braket}
\usepackage{amssymb}   
\usepackage{epstopdf}
\usepackage{xfrac}
\usepackage{subfigure}
\usepackage{anyfontsize}

\usepackage{amsmath}
\usepackage{amsfonts}
\usepackage{bbm}
\definecolor{darkblue}{rgb}{0.1,0.2,0.6}
\definecolor{darkred}{rgb}{0.8,0.1,0.2}
\usepackage[colorlinks,citecolor=darkblue,linkcolor=darkred,urlcolor=darkblue]{hyperref} 
\usepackage{tikz}
\usepackage{enumerate}
\usepackage{setspace}
\usepackage{url}  
\usepackage{mathrsfs}

\newcommand{\Tr}{\text{Tr}}

\newcommand{\tr}{\text{Tr}}

\usepackage[normalem]{ulem}

\begin{document}
\title{Partial time-reversal transformation and entanglement negativity in fermionic systems}
\author{Hassan Shapourian}
\author{Ken Shiozaki}
\author{Shinsei Ryu}
\address{Department of Physics, University of Illinois at Urbana-Champaign, Urbana Illinois 61801, USA}
\date{\today}

\begin{abstract}
{\it The partial transpose} of density matrices in many-body quantum systems, 
in which one takes the transpose only for a subsystem of the full Hilbert space,   
has been recognized as a useful tool to diagnose quantum entanglement. 
It can be used, for example, to define the (logarithmic) negativity.
For fermionic systems, 
it has been known that the partial transpose of Gaussian fermionic density matrices
is not Gaussian.
In this work,
we propose to use {\it partial time-reversal} transformation
to define (an analog of) the entanglement negativity and related quantities.  
We demonstrate that for the symmetry-protected topological phase realized in the Kitaev chain 
the conventional definition of the partial transpose (and hence 
the entanglement negativity) fails to capture the formation of the edge Majorana fermions, 
while the partial time-reversal 
computes 
the quantum dimension of the
Majorana fermions.
Furthermore, we show that
the partial time-reversal of fermionic density matrices 
is Gaussian and can be computed efficiently.
Various results (both numerical and analytical) 
for 
the entanglement negativity using the partial-time reversal 
are presented 
for 
(1+1)-dimensional conformal field theories, 
and also for fermionic disordered systems (random single phases). 
\end{abstract}

\maketitle


\section{Introduction}

Entanglement has been perceived as one of the intriguing manifestations of quantum phenomena, since the birth of quantum mechanics. There have been numerous attempts in developing theoretical tools to detect the quantum entanglement (see for example the review articles~\cite{Amico_rev2008,Calabrese_intro2009}). For a pure state, the entanglement between two complementary subsystems can be captured by the von Neumann entanglement entropy, in which case, it may assume an area law for gapped phases~\cite{Plenio_rev2010} or a universal logarithmic dependence for critical phases~\cite{Vidal2003,Calabrese2004,Calabrese2009}.
One important application of the bipartite entanglement entropy is the identification and characterization of topological phases of matter through the subleading topological term in the entanglement entropy~\cite{Kitaev_Preskill2006,Levin_Wen2006}.
For a mixed state, however, the situation is more complicated and rather open. In this case, the most practical method to capture the entanglement in extended systems is considered to be the (logarithmic) negativity~\cite{PlenioEisert1999,Vidal2002,Plenio2005} of partially transposed density matrix, inspired by the separability criterion based on the negative eigenvalues of partial transpose~\cite{Peres1996,Horodecki1996}. In parallel, the condition of separability from quantum information point of view was also studied in bosonic~\cite{Simon2000,PhysRevLett.86.3658,PhysRevLett.87.167904,RevModPhys.84.621} and fermionic~\cite{PhysRevA.76.022311,Baculs2009,Botero200439,Benatti2012,Benatti2014} systems.

Different approaches have been devised to efficiently compute the logarithmic negativity in various systems. 
Harmonic oscillator chains were studied using the covariance matrix technique~\cite{PhysRevA.66.042327,PhysRevLett.100.080502,PhysRevA.78.012335,Anders2008,PhysRevA.77.062102,PhysRevA.80.012325} 
and quantum spin chains were studied using the density matrix renormalization group \cite{PhysRevA.80.010304,PhysRevLett.105.187204,PhysRevB.81.064429,PhysRevLett.109.066403,Ruggiero_1}
and exactly~\cite{PhysRevA.81.032311,PhysRevA.84.062307}. The topologically ordered phases were also investigated for the (2+1) dimensional Chern-Simons theories~\cite{Wen2016_1,Wen2016_2} and for the toric code where exact calculations are applicable~\cite{PhysRevA.88.042319,PhysRevA.88.042318}. 
A particularly important progress was due to a systematic approach developed for 
conformal field theories (CFTs) 
\cite{Calabrese2012,Calabrese2013}. 
This approach was further expanded to study 
massive quantum field theories~\cite{Blondeau2016} as well as finite-temperature \cite{Calabrese_Ft2015,Eisler_Neq,Sherman2016}
and out-of-equilibrium situations \cite{Eisler_Neq,Coser_quench2014,Hoogeveen2015, PhysRevB.92.075109}. The entanglement spectrum of partial transpose in CFTs was also recently studied \cite{Ruggiero_2}.
Among other useful numerical methods, tree tensor network \cite{Clabrese_network2013}, Monte Carlo simulations \cite{Alba2013,PhysRevB.90.064401}, and rational interpolations \cite{Nobili2015}
are notable.

In this work, we would like to revisit the problem of finding the entanglement in fermionic mixed states.
In the previous studies on fermionic systems, initial work by Eisler and Zimbor\'as~\cite{Eisler2015} and later works~\cite{Coser2015_1,Coser2016_1,Coser2016_2,PhysRevB.93.115148,PoYao2016,Herzog2016},
the definition of the fermionic partial transpose was based on the partial transpose of the corresponding bosonic density matrix.
In other words, the action of the partial transpose on a fermionic density matrix is obtained from the action 
of the partial transpose on the bosonic density matrix through a Jordan-Wigner transformation. 
Using this definition, it was observed that for fermionic Gaussian states (free fermions), the partially transposed density matrix can be written as a linear combination of two Gaussian operators. 
In general, these two Gaussian operators do not commute and therefore, the entanglement spectrum of partial transpose cannot be easily found even for noninteracting fermions as is done for the original (untransposed) Gaussian density matrix in terms of covariance matrices~\cite{Peschel_Eisler2009}.  
This observation is then regarded as a technical difficulty for fermions~\cite{Eisler2015}, compared to bosonic systems where the partial transpose of a Gaussian density matrix is itself Gaussian~\cite{PhysRevA.66.042327,PhysRevA.71.032349}.

Here, we present a different definition of fermionic partial transpose, in contrast to the prior definition of fermionic partial transpose~\cite{Eisler2015}, 
based on the partial time-reversal transformation.
Throughout this paper, we use the term ``partial time-reversal (TR)'' to refer to our definition and reserve the term ``partial transpose" for the original definition in Ref.~\cite{Eisler2015}.
This is motivated by our earlier observation 
that the partial time-reversal can be used as a means to detect time-reversal symmetry protected topological (SPT) phases of fermions or bosons~\cite{Shap2016,Shiozaki2016}. 
More specifically, it can be used to construct 
topological invariants of SPT phases protected by time-reversal.
These topological invariants are intimately connected to the partition functions
(path integrals) of {\it spin} topological quantum field theories. 
(Here, spin topological quantum field theories are effective field theories
that can describe gapped phases of fermionic systems in the limit of zero
correlation length, as opposed to ordinary topological quantum field theories,
for which we do not consider fundamental fermonic degrees of freedom.)
For example, in (1+1)-dimensional SPT phases protected by time-reversal,
their topological invariants are given by the partition functions 
on unoriented spacetime manifolds such as the real-projective plane $\mathbb{R}P^2$ or the Klein bottle ~\cite{2014arXiv1403.1467K,*Kapustin2b,*Kapustin2015b,*Kapustin2015f,Freed2014,*Freed2016}.
Using the insights from spin topological quantum field theories, 
our definition of the partial TR is such that it can be used to 
reproduce the results from spin topological quantum field theories
(at least in the limit of zero correlation length).


Compared with the partial transpose, our definition has two useful properties: 
First of all, the partial TR can capture the Majorana bonds between the two subsystems, while the partial transpose cannot.
We demonstrate this for the symmetry-protected topological phase realized in the Kitaev chain.
(See Sec.\ \ref{sec:examples} and Fig.\ \ref{fig:Kitaev_vs_mu}.) 
Second, the partial TR of a fermionic Gaussian state remains Gaussian. 
As we have shown in Appendix~\ref{sec:Eisler}, the partial TR is equal to one of the Gaussian operators already obtained in the partial transpose.

We should also note that unlike the bosonic systems, 
the
partial transpose for fermions may be accompanied by a sign structure (or more generally a complex phase) as fermion operators obey anti-commuting algebra. In this respect, the stark difference between 
the
partial TR and the partial transpose is that 
the
partial TR, $\rho^R$, in general is not Hermitian and its eigenvalues might be complex. However, one can still define an entanglement measure by considering the eigenvalues of $\sqrt{\rho^R \rho^{R\dag}}$.

%
Our paper is organized as follows: In Sec.~\ref{sec:partial_def}, 
we briefly review the definition of the bosonic partial transpose and its relation to the time-reversal transformation, then we define the fermionic partial time-reversal and write the corresponding transformation rules for the fermionic density matrix in the occupation number basis and in the operator formalism in terms of Majorana operators. In Sec.~\ref{sec:neg_proc}, we provide a general procedure to obtain the spectrum of 
the partial TR for noninteracting fermions and present a path integral picture for 
the partial TR using the replica trick.
In Sec.~\ref{sec:examples}, we study two examples of the Kitaev chain and the Su-Schrieffer-Heeger chain and compare the resulting entanglement computed by our definition and the previous definition of the partial transpose. Using the replica approach introduced in Sec.~\ref{sec:neg_proc},  we write Fisher-Hartwig type expressions for the entanglement measure associated with the partial TR at the criticality and derive analytical formulas in Sec.~\ref{sec:CFT}. 
We also show that analytical results agree with the numerical calculations. 
Towards the end of Sec.~\ref{sec:CFT}, we numerically study the random singlet phase of the disordered XX chain and compare our results with the literature.
Finally, we finish our discussion with closing remarks in Sec.~\ref{sec:discussion}. We explain some details of our calculations in four appendices.

\section{\label{sec:partial_def} Partial transpose and partial time-reversal transformation}

In this section, we first review the definition of the partial transpose for a bosonic density matrix and its relation to the partial time-reversal transformation. Next, we introduce a definition of partial TR for fermions.

Our focus in this paper is on the tripartite geometry. Consider a subsystem $A=A_1\cup A_2$ which is in turn partitioned into two smaller subsystems $A_1$ and $A_2$ and its reduced density matrix $\rho_A=\text{tr}_{B}(\rho)$ after tracing out the rest of the system $B$.
The  logarithmic negativity is defined by
\begin{align} \label{eq:neg_def}
{\cal E}:= \ln \Tr |\rho_A^{T_1}|, 
\end{align}
where $\rho_A^{T_1}$ denotes the partial transpose of the density matrix and $\Tr |O|:=\Tr \sqrt{O^\dag O}$ represents the trace norm of the operator $O$, which is the sum of the square roots of the eigenvalues of the product operator $O^\dag O$. When $O$ is Hermitian, the trace norm is simplified into the absolute value of the eigenvalues of $O$.  Now, let us discuss the partial transpose.

\subsection{Bosons}
A generic bosonic density matrix can be written in the following form
\begin{align}
\rho_A = \sum_{ijkl} \braket{e_i^1,e_j^2|\rho_A |e_k^1,e_l^2} \ket{e_i^1,e_j^2} \bra{e_k^1,e_l^2}, 
\end{align}
where $\ket{e_j^{1}}$ and $\ket{e_k^{2}}$ denote orthonormal bases in the Hilbert spaces $\mathcal{H}_1$ and $\mathcal{H}_2$ corresponding to the $A_1$ and $A_2$ regions. 
The partial transpose of a density matrix for the subsystem $A_1$ is defined by exchanging the matrix elements in the subsystem $A_1$ as in
\begin{align}
\label{eq:parttrans_b}
\rho_A^{T_1} := \sum_{ijkl} \braket{e_k^1,e_j^2|\rho_A |e_i^1,e_l^2} \ket{e_i^1,e_j^2} \bra{e_k^1,e_l^2}, 
\end{align}
which is equivalent to the following transformation in the operator basis 
\begin{align}
\big( \ket{e_i^1,e_j^2} \bra{e_k^1,e_l^2} \big)^{T_1} := \ket{e_k^1,e_j^2} \bra{e_i^1,e_l^2}. 
\label{eq:app_f1}
\end{align}

As was shown by Simon~\cite{Simon2000}, the partial transpose has a geometric interpretation as partial TR or mirror reflection in phase space. This idea can be readily illustrated in a single bosonic mode defined in terms of the operators ${a}=({q}+i{p})/\sqrt{2}$ and ${a}^\dag=({q}-i{p})/\sqrt{2}$ where ${q}$ and ${p}$ are position and momentum operators, respectively, which obey the canonical commutation relation $[{q},{p}]=i$ or equivalently $[a,a^\dag]=1$. The TR transformation for a basis vector $\ket{\alpha}\bra{\alpha^\ast}$ in the coherent state representation~\cite{Wigner1932} is given by
\begin{align}
\label{eq:coh_b}
  \ket{\alpha} \bra{\alpha^\ast} 
 \to \ket{\alpha^\ast} \bra{\alpha} :=\left(\ket{\alpha} \bra{\alpha^\ast} \right)^T 
\end{align}
in which $\ket{\alpha} = e^{\alpha a^\dag}\ket{0}$ and $\bra{\alpha^\ast} = \bra{0} e^{ \alpha^\ast a}$ are coherent states and $\alpha$ and $\alpha^\ast$ are complex numbers and we use the fact that the TR operator  is simply equal to complex conjugation, ${\cal T}={\cal K}$.
This definition for the partial transpose leads to the transformation rule $\ket{m}\bra{n}\mapsto \ket{n}\bra{m}$ in the occupation number basis, which follows from identifying the same monomials of $\alpha$ and $\alpha^\ast$ on both sides of the second equality in the above definition.
 It is easy to see that the associated Wigner distribution function $W(q,p)$ goes to $W(q,-p)$~\cite{Simon2000}. Hence, this means that for bosonic systems partial transpose is the same as partial time-reversal or mirror reflection in $(q,p)$-space. This fact was also used in harmonic chains to calculate the negativity in terms of the covariance matrix~\cite{PhysRevA.66.042327}.

\subsection{Fermions}
\subsubsection{Definition in coherent state basis}
It is worth noting that for a bosonic system either Eqs.~(\ref{eq:app_f1}) or (\ref{eq:coh_b}) can be used as a fundamental defining equation for the partial transpose. However, this is not the case for fermions. Due to anti-commuting property of fermions, Eq.~(\ref{eq:app_f1}) may acquire an additional minus sign depending on the states being transposed. Therefore, we propose to use the fermionic version of the TR transformation as a guiding principle to determine the rules associated with the partial transpose. To begin with, let us consider a single-site system described by fermionic operators $f$ and $f^\dag$ which obey the anti-commutation relation $\{f,f^\dag\}=1$. 
We define the analog of Eq.~(\ref{eq:coh_b}) for fermions, 
\begin{align}
\label{eq:coh_f}
 \ket{\xi} \bra{\bar{\xi}}  \to \ket{i \bar{\xi}} \bra{i \xi} := 
U \left(\ket{\xi} \bra{\bar{\xi}} \right)^R U^\dag
\end{align}
where $\xi, \bar \xi$ are Grassmann variables, $\ket{\xi} = e^{-\xi f^{\dag}}\ket{0}$ and $\bra{\bar \xi} = \bra{0} e^{-  f \bar \xi}$ are fermionic coherent states, and $U$ is the unitary part of TR operator, ${\cal T}=U {\cal K}$. We use the superscript $R$ to distinguish our definition of partial TR from the previous definition of partial transpose for fermions~\cite{Eisler2015}. 
It is important to note that the TR transformation is not just exchanging Grassmann variables between bra and ket states but also multiplying them by a factor of $i$
\footnote{More generally, one can consider $\ket{-e^{-i\theta} \bar{\xi}}\bra{e^{i\theta} {\xi}}$ for arbitrary $\theta\in [0,2\pi)$ as the definition of partial TR. All different definitions lead to the same spectrum for the partial TR, as they are related by unitary transformations. We choose $\theta=\pi/2$, since the transformation rule for Majorana operators looks simpler.}. The factor of $i$ appears due to anticommuting property of Grassmann variables and is required for keeping the trace
(i.e., sum of the diagonal elements of density matrix, $1$ and $f^\dag f$) 
unchanged after taking the transpose. Similar transformation rules of the Grassmann variables can be derived for the time-reversal transformation in the path integral formalism~\cite{Shap2016}.

It is worth noting that fermionic coherent states are Grassmann even and commute with each other. Therefore, Eq.~(\ref{eq:coh_f}) can be readily generalized for a many-particle system and the partial TR with respect to the interval $A_1$ reads as
\begin{align}
&
U_{A_1} \big( \ket{\{ \xi_j \}_{j \in A_1}, \{\xi_j\}_{j \in A_2} } \bra{ \{ \bar{\chi}_j \}_{j \in A_1}, \{\bar{\chi}_j\}_{j \in A_2} } \big)^{R_1} U_{A_1}^{\dag}
 \nonumber \\
&= \ket{\{ i \bar \chi_j \}_{j \in A_1}, \{\xi_j\}_{j \in A_2} } \bra{ \{ i \xi_j \}_{j \in A_1}, \{\bar{\chi}_j\}_{j \in A_2} },
\label{eq:app_f2}
\end{align} 
where $U_{A_1}$ acts only on the Hilbert space of $A_1$ and $\ket{\{\xi_j\}} = e^{-\sum_j \xi_j f^{\dag}_j}\ket{0}$ and $\bra{\{\bar \chi_j\}} = \bra{0} e^{- \sum_j f_j \bar \chi_j}$ are the many-particle fermionic coherent states. Consider the occupation number basis,
\begin{align} \label{eq:ptrans_f}
\ket{\{ n_j \}_{j \in A_1} , \{n_j\}_{j \in A_2}} 
&= (f_{m_1}^{\dag})^{n_{m_1}} \cdots (f_{m_{\ell_2}'}^{\dag})^{n_{m_{\ell_2}'}} \ket{0}
\end{align}
where $n_j, \bar n_j \in \{0,1\}$ for the subsystem $A=A_1 \cup A_2 $. Throughout this paper, the intervals $A_1$ and $A_2$ have lengths $\ell_1$ and $\ell_2$  and we use the indices $\{m_1, \dots, m_{\ell_1}\} \cup \{m_1', \dots, m'_{\ell_2}\}$ to denote the sites in the subsystem (adjacent or disjoint intervals). 
 The definition (\ref{eq:app_f2}) in the occupation number basis
can be viewed as a ``partial transposition'' similar to (\ref{eq:app_f1}),
\begin{align}
&
U_{A_1} \left( \ket{\{ n_j \}_{j \in A_1} , \{n_j\}_{j \in A_2}} \bra{\{ \bar n_j \}_{j \in A_1}, \{ \bar n_j \}_{j \in A_2} } \right)^{R_1} U_{A_1}^\dag
\nonumber \\
&= \ket{\{ \bar n_j \}_{j \in A_1} , \{n_j\}_{j \in A_2}} \bra{\{ n_j \}_{j \in A_1}, \{ \bar n_j \}_{j \in A_2} }\nonumber \\ & \ \ \ \times  (-1)^{\phi(\{n_j\}, \{\bar n_j\})}, 
\label{eq:app_f_21}
\end{align}
up to the phase factor $(-1)^{i \phi(\{n_j\}, \{\bar n_j\})}$  given by
\begin{align}
\phi(\{n_j\}, \{\bar n_j\}) =& \frac{\tau_1 (\tau_1+2)}{2}+ \frac{\bar{\tau}_1 (\bar{\tau}_1+2)}{2}+ \bar{\tau} _2\tau_2 
 \nonumber\\
 &+ {\tau}_1\tau_2+ \bar{\tau}_1\bar{\tau}_2+ (\bar{\tau}_1+\bar{\tau}_2)(\tau_1+\tau_2)
\end{align}
in which
$\tau_s=\sum_{j\in A_s} n_j$  and $\bar{\tau}_s=\sum_{j\in A_s} \bar{n}_j$, are the number of occupied states in the $A_s$ interval.

\subsubsection{Defintion in terms of Majorana operators}
Let us now explain the partial TR in the operator formalism and in particular the unitary operator $U_{A_1}$.
We introduce the Majorana operators by 
\begin{align} \label{eq:Majorana}
c_{2j-1} :=f_j+f_j^\dag, \quad c_{2j}:= i (f_j-f_j^\dag).
\end{align}
Our goal here is to determine how Majorana operators transform under the transformation rule defined by (\ref{eq:app_f2}).  This gives us an idea about what this transformation does compared to the previous definition for the partial transpose~\cite{Eisler2015}.
Recall that the density matrix in the Majorana representation can be written as 
\begin{align} \label{eq:denmat}
\rho_A 
= \sum_{\substack{\kappa, \tau,\\ |\kappa|+|\tau| = {\rm even}}} w_{\kappa, \tau} 
c^{\kappa_{m_1}}_{{m_1}} \cdots c^{\kappa_{2m_{\ell_1}}}_{2m_{\ell_1}} 
c^{\tau_{m'_1}}_{m'_1} \cdots c^{\tau_{2m'_{\ell_2}}}_{2m_{\ell_2}'} .
\end{align}
We use the notation $c_x^0=\mathbb{I}$ and $c_x^1=c_x$, so that $\kappa_i, \tau_j \in \{0,1\}$ and also introduce the vectors (bit strings) $\kappa = (\kappa_{m_1}, \dots \kappa_{2m_{\ell_1}})$, 
$\tau = (\tau_{m_1'}, \dots, \tau_{2m_{\ell_2}'})$, 
and their norms $|\kappa| = \sum_j \kappa_j, |\tau| = \sum_j \tau_j$. 
We note  that in the fermionic systems with global fermion-number parity symmetry the reduced density matrix of subsystem is always parity-even~\cite{PhysRevA.76.022311,Baculs2009,Zimboras2014}, which implies that the sum is over all possible bit strings, provided that $|\kappa|+|\tau| = {\rm even}$. 
Using the above expression for the density matrix, the partial TR with respect to the subsystem $A_1$ is defined by
\begin{align*}
\rho^{R_1}_A 
:= \sum_{\substack{\kappa, \tau,\\ |\kappa|+|\tau| = {\rm even}}} w_{\kappa, \tau} 
\mathscr{R}(c^{\kappa_{m_1}}_{{m_1}} \cdots c^{\kappa_{2m_{\ell_1}}}_{2m_{\ell_1}} )
c^{\tau_{m'_1}}_{m'_1} \cdots c^{\tau_{2m'_{\ell_2}}}_{2m_{\ell_2}'} .
\end{align*}
In order to determine the transformation rules, we should note that the transformation in (\ref{eq:app_f2}) consists of two steps: Taking the partial transpose $\rho_A^{R_1}$ and applying the unitary transformation $U_{A_1} \rho_A^{R_1} U_{A_1}^\dag$. In what follows, we will show what each step does to Majorana operators.

First, we need to introduce the unitary part $U_{A_1}$ of the TR operator which acts only on the subsystem $A_1$. 
We demand that the TR operator ${\cal T}$ must satisfy 
\begin{align}
{\cal T} f_j {\cal T}^{-1} = f_j, \qquad {\cal T}i {\cal T}^{-1}=-i,
\end{align}
for spinless fermions. The definition of Majorana operators in (\ref{eq:Majorana}) then implies that ${\cal T} = (\prod_j c_{2j-1}) {\cal K}$ where ${\cal K}$ is the complex conjugation. 
Therefore, we define $U_{A_1}$ as a unitary part acting on the subsystem $A_1$:
\begin{align}
U_{A_1} = \prod_{j \in A_1} c_{2j-1}. 
\label{eq:app_f_ut_part}
\end{align}
Second, the action of $\mathscr{R}$ on Majorana operators is uniquely determined by identifying the same monomials in terms of Grassmann variables on both sides of (\ref{eq:app_f2}). These identifications fix the partial transpose $\mathscr{R}(c) = i c$  and enforces the condition $\mathscr{R}(M_1 M_2) = \mathscr{R}(M_1) \mathscr{R}(M_2)$ for two arbitrary operators acting on the Hilbert space of $A_1$ (Details of this derivation can be found in \cite{Shap2016}). This means that there is no reversed ordering in our transformation rule which contrasts with Ref.~\cite{Eisler2015}. 

To sum up, we find that the partial TR is given by 
\begin{align}
\label{eq:ptrans_Maj}
\rho^{R_1}_A
= \sum_{\substack{\kappa, \tau,\\ |\kappa|+|\tau| = {\rm even}}} w_{\kappa, \tau} 
i^{|\kappa| }  c^{\kappa_{m_1}}_{{m_1}} \cdots c^{\kappa_{2m_{\ell_1}}}_{2m_{\ell_1}} 
c^{\tau_{m'_1}}_{m'_1} \cdots c^{\tau_{2m'_{\ell_2}}}_{2m_{\ell_2}'} .
\end{align}
It is straightforward to check that the above definition fulfills three natural expectations for a partial transpose:
(i) subsequent partial transpositions for two intervals are identified with full transposition 
\begin{align}
(\rho_A^{R_1})^{R_2} = \rho_A^R, 
\end{align}
 (ii) applying full transposition twice returns the original density matrix 
\begin{align}
 (\rho_A^R)^R = \rho_A, 
\end{align}
 and (iii) when considering $n$ flavors of fermions 
\begin{align}
 (\rho_{1,A } \otimes\cdots \otimes \rho_{n,A })^{R_1}= (\rho_{1,A })^{R_1}  \otimes\cdots \otimes (\rho_{n,A })^{R_1}.
\end{align}
 As we show in Appendix~\ref{sec:Eisler}, the partial transpose originally proposed in Ref.~\cite{Eisler2015} does not satisfy the conditions (i) and (iii). 
 Condition (iii) is particularly important and is necessary, since the logarithmic negativity (as a measure of entanglement) is required to be additive,~i.e.,
 \begin{align} \label{eq:additive}
 {\cal E}(\rho_{1,A } \otimes\cdots \otimes \rho_{n,A })=  {\cal E}(\rho_{1,A })+ \cdots+  {\cal E}(\rho_{n,A }).
 \end{align}
 However, we observe that the partial transpose does not commute with taking the tensor products and hence, it does not fulfill the condition of additivity (\ref{eq:additive}). In Appendix~\ref{sec:Eisler}, we show that the violation of additivity let the partial transpose capture the singlet bonds while it cannot capture the Majorana bonds.

We should note that the matrix resulting from the partial TR of the density matrix is not Hermitian and may have complex eigenvalues. Nevertheless, we can still use Eq.~(\ref{eq:neg_def}) to define the logarithmic negativity where the eigenvalues of the combined operator $[ \rho_A^{R_1}\rho_A^{R_1\dag}]$ are real-valued.
In the rest of the paper, we shall use the term ``logarithmic negativity'' to refer to the quantity 
\begin{align} \label{eq:neg_pTR}
{\cal E} := \ln \Tr|\rho_A^{R_1}|=  \ln \Tr \sqrt{\rho_A^{R_1}(\rho_A^{R_1})^\dag}
\end{align}
although the original intuition behind the naming of the negativity, 
as a measure of how negative the eigenvalues are, bears no meaning here.

\section{\label{sec:neg_proc} General procedure for computing the entanglement negativity}

In this section, we present two methods to compute the moments of the partially transformed density matrix $\rho^{R_1}_A$ and ultimately the logarithmic negativity. 
 In the first part, we develop a method to calculate the entanglement negativity for noninteracting fermions
 using the coherent state representation.
In the second part, we derive a general way to construct the moments of transposed density matrix using the replica approach \cite{Calabrese2012,Calabrese2013,Herzog2016} and
provide an equivalent spacetime picture of this quantity. The result of this construction can be recast in the form of ground state expectation values of partial $U(1)$ phase twist operators  $\exp[{i\theta_i \sum_{j\in A_i} f^\dag_j f_j}]$ which act on the subsystems $A_1$ and $A_2$ with different twist angles $\theta_i$.  This treatment is quite general and can be applied to many-particle wave functions as well.
 Later in Sec.~\ref{sec:CFT}, we will use this method to analytically derive the negativity at the critical points.

It is important to note that the partial TR in either forms $\rho_A^{R_1}$ or $U_{A_1}\rho_A^{R_1}U_{A_1}^{\dag}$ are related via the unitary operator $U_{A_1}$ and as far as the negativity $\Tr|\rho_A^{R_1}|$ or the moments of transposed density matrix $\Tr(\rho_A^{R_1}\rho_A^{R_1\dag}
\rho_A^{R_1}\rho_A^{R_1\dag}\cdots)$ is concerned, the result is the same. 
So, from now on, we use them interchangeably.


\subsection{Noninteracting fermions}

Here, we discuss how to compute the entanglement negativity for noninteracting fermions. 
In particular, we show that the transformed operator $\rho_A^{R_1}$ can be written in the Gaussian form (exponentiated bilinear) similar to the original density matrix. Therefore, one can simply compute the eigenvalues of the partially TR transformed density matrix and obtain the logarithmic negativity.

We consider a general form of quadratic Hamiltonians,
\begin{align}
\hat{H}= \sum_{i,j}  t_{ij} f^\dag_i f_j +\Delta_{ij} f^\dag_i f_j^\dag + \text{H.c.}
\end{align}
The reduced density matrix of such Hamiltonians can also be recast in a quadratic form 
\begin{align}  \label{eq:red_den}
\rho_A =\frac{e^{-\hat{\cal H}_I}}{\cal Z}
\end{align}
where the entanglement Hamiltonian is $\hat{\cal H}_I= \sum_{i,j} {h}^1_{ij} f_i^\dag f_j+ {h}^2_{ij} f_i^\dag f_j^\dag + \text{H.c.}$ and ${\cal Z}$ is the normalization factor. The eigenvalues of $\hat{\cal H}_I$ can be found in terms of generalized Green function which includes the pairing correlators~\cite{Peschel_Eisler2009,Borchmann2014},
\begin{align} \label{eq:Gmat}
G_{ij}=\left(\begin{array}{cc}
1-[C^T]_{ij} & [F^\dag]_{ij} \\
F_{ij} & C_{ij}
\end{array}
\right),
\end{align}
where
\begin{align} \label{eq:corr}
C_{ij}=\braket{f_i^\dag f_j}, \qquad
F_{ij}=\braket{f_i^\dag f_j^\dag},\end{align}
are the ground state two-body correlators and the particle-hole correlators, respectively. By definition, they satisfy $C^\dag=C$ and $F^T=-F$.
The eigenvalues of the $G$ matrix can be recast in the form of pairs $(\alpha_i,1-\alpha_i)$. It is well-known that $G$ and $\hat{\cal H}_I$ can be simultaneously diagonalized and the eigenvalues of $G$ are related to those of $\hat{\cal H}_I$ (denoted by $\zeta_i$) through $\zeta_i = \ln\left(\frac{1-\alpha_i}{\alpha_i}\right)$. Given the eigenvalues of $\hat{\cal H}_I$, one can easily compute various entanglement measures.

Let us now discuss how to construct the partial TR in the coherent state representation. The reduced density matrix (\ref{eq:red_den}) can be represented in the coherent state basis by
\begin{align}
\rho_A &=\frac{1}{{\cal Z}_\rho} \int {d}[ \xi] {d} [\bar{\xi}]  \ e^{\frac{1}{2} \sum_{i,j \in A} \boldsymbol{\xi}_i^T S_{ij} \boldsymbol{\xi}_j }  
  \ket{\{ \xi_j \}_{j \in A} } \bra{ \{ \bar{\xi}_j \}_{j \in A}} 
\end{align}
where the $S_{ij}$ matrix is given by
\begin{align}
S_{ij}= \Gamma_{ij} + i\sigma_2\ \delta_{ij},
\end{align}
 the matrix $\Gamma$ is related to the Green function matrix (\ref{eq:Gmat}) through
\begin{align} \label{eq:Gammadef}
[\Gamma^{-1}]_{ij}=\left( \begin{array}{cc}
[F^\dag]_{ij} & \textcolor{red}{+}[C^T]_{ij} \\
\textcolor{red}{-}C_{ij} & F_{ij}
\end{array} \right),
\end{align}
the Grassmannian vector is $\boldsymbol{\xi}_j^T= (\xi_j,\bar{\xi}_j)$ in the particle-hole basis as introduced above and ${\cal Z}_\rho=\text{Pf}[S-i\sigma_2]=\text{Pf}[\Gamma]$.
Here, $\sigma_2$ is the Pauli matrix in the particle-hole basis and appears as the normalization factor $e^{-\bar{\xi}\xi} $ in $e^{-\bar{\xi}\xi} \ket{\xi}\bra{\bar{\xi}}$.  This representation of density matrix manifestly yields the identities in (\ref{eq:corr}) and the fact that all higher-order correlators can be computed using the Wick expansion.

Using (\ref{eq:app_f2}), we can write the partial TR transformation as
\begin{align}
&\rho_A^{R_1}   =\frac{1}{{\cal Z}_\rho} \int {d}[ \xi] {d} [\bar{\xi}]  \ e^{\frac{1}{2} \sum_{i,j\in A}\boldsymbol{\xi}_i^T S_{ij} \boldsymbol{\xi}_j } \nonumber \\
& \times 
\ket{\{i \bar{\xi}_j \}_{j \in A_1}, \{\xi_j\}_{j \in A_2} } \bra{ \{ i{\xi}_j \}_{j \in A_1}, \{\bar{\xi}_j\}_{j \in A_2} } 
\end{align}
This transformation can be absorbed into a redefinition of the $S_{ij}$ matrix after introducing the new variables  $\boldsymbol{\xi}= U_S \boldsymbol{\chi}$,
\begin{align}
\rho_A^{R_1}  =& \frac{1}{{\cal Z}_\rho} \int {d}[ \chi] {d} [\bar{\chi}]  \ e^{\frac{1}{2} \sum_{i,j\in A} \boldsymbol{\chi}_i^T {S}^{R_1}_{ij} \boldsymbol{\chi}_j }  
\ket{\{ \chi_j \}_{j \in A} } \bra{ \{ \bar{\chi}_j \}_{j \in A}} 
\end{align}
where $S^{R_1}=U_{S}^T S U_{S}$ and $U_{S}=U_{S}^T$ is a permutation matrix 
\begin{align} \label{eq:Umat}
U_S= \left(\begin{array}{cccc}
0 & 0 & -i \mathbb{I}_{11} & 0 \\
0 & \mathbb{I}_{22} & 0 & 0\\
-i \mathbb{I}_{11} & 0 & 0  & 0 \\
0 & 0 & 0 & \mathbb{I}_{22}
\end{array}\right),
\end{align}
in the $(\{\xi_j\}_{j \in A_1},\{\xi_j\}_{j \in A_2},\{\bar{\xi}_j\}_{j \in A_1},\{\bar{\xi}_j\}_{j \in A_2})$ basis. Here, $\mathbb{I}_{11}$ and $\mathbb{I}_{22}$ are identity matrices acting on $A_1$ and $A_2$ subsystems, respectively. As a result of this transformation, we get the new matrix $\Gamma^{R_1}$,
\begin{align} \label{eq:Gammatilde}
[{\Gamma}^{R_1}]^{-1}&=
(S^{R_1}-i\sigma_2)^{-1}=
\left( \begin{array}{cc}
F'^{R_1} & \textcolor{red}{+}[C^{R_1}]^T \\
\textcolor{red}{-}C^{R_1} & {F^{R_1}}
\end{array} \right),
\end{align}
which yields the transformed correlators ${F}^{R_1}$, ${F}'^{R_1}$ and ${C}^{R_1}$.
Generically, $\text{Re}[{F}^{R_1}]=-\text{Re}[{F}'^{R_1}]$, while their imaginary parts are not in general related and only share the same eigenvalue spectrum.

In order to obtain the negativity~(\ref{eq:neg_pTR}) of $\rho_A^{R_1}$, we need to find the eigenvalues of
the composite operator $\Xi=\rho_A^{R_1} \rho_A^{R_1\dag}$
which is also Gaussian, since the product of two Gaussian states remain Gaussian. In Appendix~\ref{sec:OpOm}, we provide details of how to construct this operator. The result is
\begin{align} \label{eq:OpOm}
\Xi =& \frac{1}{{\cal Z}_\rho^2} \int {d}[ \chi] {d} [\bar{\chi}]  \ e^{\frac{1}{2} \sum_{i,j\in A} \boldsymbol{\chi}_i^T \tilde{S}_{ij} \boldsymbol{\chi}_j } \ket{\{ \chi_j \}_{j \in A} } \bra{ \{ \bar{\chi}_j \}_{j \in A}} 
\end{align}
Therefore, we determine the reconstructed Green function $\tilde{G}$ associated with $\tilde{S}$ as in
\begin{align} 
\tilde{G}&= 
\left( \begin{array}{cc}
1-\tilde{C}^T & \tilde{F}^\dag \\
\tilde{F} & \tilde{C}
\end{array} \right),
\end{align}
where we read off $\tilde{F}$ and $\tilde{C}$ from analog of Eq.~(\ref{eq:Gammatilde}) for $\tilde{S}$.
The $2M=2(\ell_1+\ell_2)$ eigenvalues of $\tilde{G}$ are in the form of pairs $(\lambda_i,1-\lambda_i)$ where $0\leq \lambda_i\leq1$.  Hence, the logarithmic negativity can be easily computed by
\begin{align} \label{eq:coh_neg}
{\cal E} &= \ln \left[ \frac{\Tr [\sqrt{\rho_A^{R_1} \rho_A^{R_1\dag}}]}{\sqrt{\Tr [\rho_A^{R_1} \rho_A^{R_1\dag}}]} \sqrt{\Tr [\rho_A^{R_1} \rho_A^{R_1\dag}}]  \right] 
\nonumber \\ 
&= \sum_{i=1}^{M} \ln  
\left( \sqrt{\lambda_i} + \sqrt{1-\lambda_i} \right) 
+\frac{1}{2} \ln \left| \frac{\text{Pf}\, [\tilde{S}-i\sigma_2]}{\text{Pf}\, [{S}-i\sigma_2]^2} \right|,
\end{align}
where we use the identity $\Tr\, \Xi=\text{Pf}\, [\tilde{S}-i\sigma_2]/{\cal Z}_\rho^2$ by performing the Gaussian integral in~(\ref{eq:OpOm}).
We can also determine the moments of partial TR,
\begin{align} \label{eq:coh_moment}
\Tr \ \Xi^n &=  
\left[ \prod_{i=1}^{M} 
\left( 
\lambda_i^n+ (1-\lambda_i)^n \right) \right]
\left(\frac{\text{Pf}\, [\tilde{S}-i\sigma_2]}{\text{Pf}\, [{S}-i\sigma_2]^2}\right)^n.
\end{align}

At this stage, let us make some remarks about the relation between partial TR and partial transpose introduced in~\cite{Eisler2015}.
Let $P_1$ be 
the fermion number parity operator 
for subsystem $A_1$, which can be written as $P_1=\prod_{j\in_{A_1}}ic_{2j-1}c_{2j} $  in the Majorana basis (\ref{eq:Majorana}), and define the operators
$\rho_{+}=\case{1}{2}(\rho_A + P_1 \rho_A P_1)$ and 
$\rho_{-}=\case{1}{2}(\rho_A - P_1 \rho_A P_1)$. In Ref.~\cite{Eisler2015}, the partial transpose is defined such that
\begin{equation}
\rho^{T_1}_A= \frac{1}{\sqrt{2}} (e^{-i\frac{\pi}{4}} O_+ +e^{i\frac{\pi}{4}} O_- ) \, ,
\label{ptrho}
\end{equation}
where $O_+$ and $O_-$ are two Gaussian operators. As we show in Appendix~\ref{sec:Eisler}, we have the relations
\begin{align}
\rho^{R_1}_A= O_+, \qquad \rho^{R_1\dag}_A= O_-,
\end{align}
and therefore, the negativity associated with partial TR is ${\cal E}= \ln \Tr\sqrt{O_+ O_-}$. For pure states, 
it was shown by Ref.~\cite{Eisler2015} that $[O_+,O_-]=0$ and hence, our definition of the negativity is simplified into ${\cal E} = \ln \Tr |O_+|$ which coincides with that of Ref.~\cite{Eisler2015}. This is in turn identical to the $1/2$-R\'enyi entropy $S_{1/2}=2\ln\Tr (\rho_A^{1/2})$.


\subsection{\label{sec:rep} Replica approach}

Here, we present a spacetime view of moments of partial TR, $\Tr(\rho_A^{R_1}\rho_A^{R_1\dag}
\rho_A^{R_1}\rho_A^{R_1\dag}\cdots)$, and write them in terms of partition functions. We also show how this view can be adapted to  an expectation value of the ground state wave function.

\begin{figure}
\includegraphics[scale=.55]{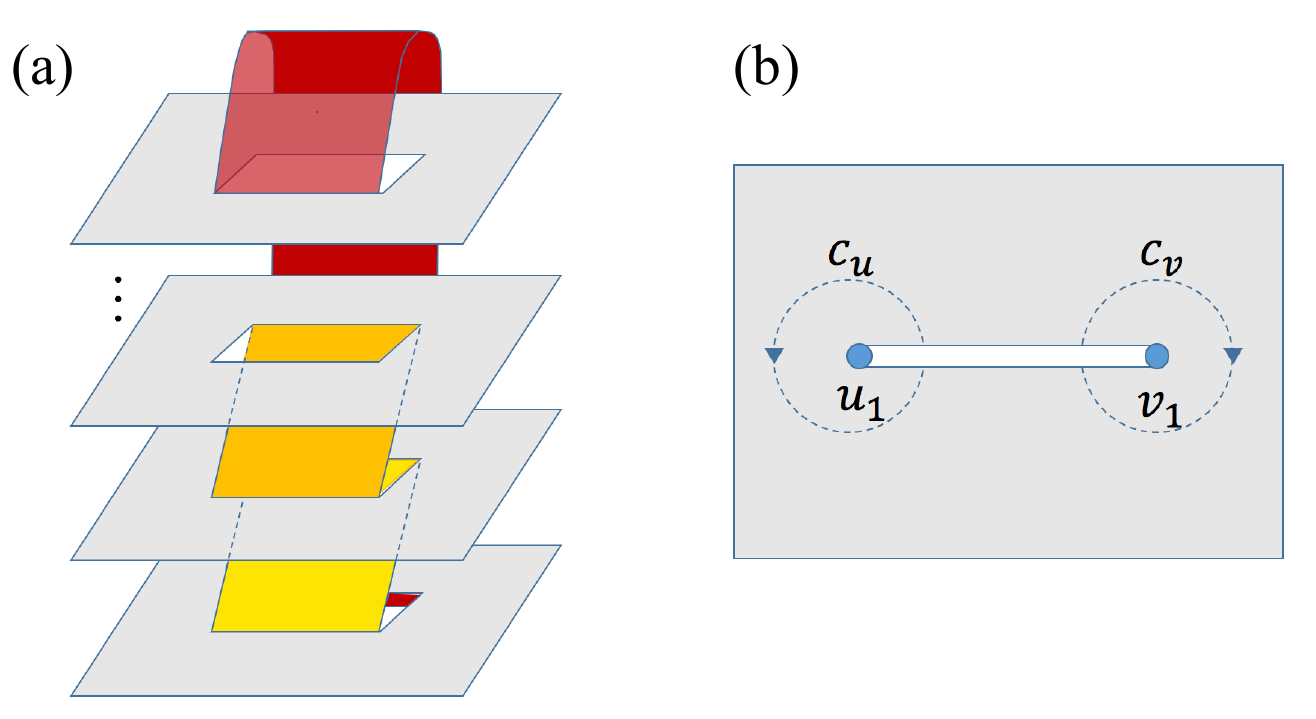}
\caption{\label{fig:repRenyi} Path integral representation of R\'enyi entropy for (a) multi-sheet and (b)  single-sheet spacetime manifold. }
\end{figure}

Before we proceed, let us briefly review the replica approach to find the regular entanglement entropy. Next, we make connections to our construction of partial TR.
The R\'enyi entanglement entropy (REE) of a reduced density matrix $\rho_A$ is defined by 
\begin{align}
S_n := \frac{1}{1-n} \ln\text{Tr} [\rho_A^n].
\end{align}
The above expression can be viewed as making cuts in the spacetime manifolds of $n$ flavors (replicas) $\psi_i$ and glue them in order along the cuts, as shown in Fig.~\ref{fig:repRenyi}(a). 
 Alternatively, one can consider multi-value field $\Psi$,
 \begin{align}
\Psi= (\psi_1,\cdots,\psi_n)^T,
\end{align}
 on a single-sheet spacetime as in Fig.~\ref{fig:repRenyi}(b). This way when we traverse a circuit around $u_1$ or $v_1$  (denoted as $C_u$ and $C_v$ in Fig.~\ref{fig:repRenyi}(b)) the fermion operator transforms as
\begin{align} \label{eq:Ttrans}
{\psi}_i \mapsto T_{ij} \psi_j
\end{align}
where the twist matrix $T$ is given by
\begin{align} \label{eq:Tmat}
T=\left(\begin{array}{cccc}
0 & -1 & 0 & \dots \\
0 & 0 & -1 & 0  \\
\vdots & \vdots & \ddots & -1    \\
1 & 0 & \cdots & 0 \\
\end{array} \right).
\end{align}
The idea of
using the twist matrix for 
the REE of the Dirac fermions was originally initiated by 
Casini~{\it et.~al.}~\cite{Casini2005} 
where 
they proposed the
twist matrix is unitarily equivalent to (\ref{eq:Tmat}). 
The fermionic twist matrix approach was further applied to the entanglement negativity of disjoint intervals~\cite{Herzog2016}.
Here, we present an explicit derivation of the twist matrix using the coherent state representation. A generic density matrix can be represented in the fermionic coherent-state as
\begin{align}
\rho_A = \int d\alpha d\bar\alpha\ d\beta d\bar\beta\ \rho_A(\bar\alpha,\beta) \ket{\alpha}\bra{\bar\beta} e^{-\bar\alpha \alpha- \bar\beta\beta} 
\end{align}
where $\alpha$, $\bar\alpha$, $\beta$ and $\bar\beta$ are independent Grassmann variables and we did not show the real-space (and possibly other) indices for simplicity. We also absorb the normalization factor into the integral measure.
The trace formula then reads
\begin{align}
\tr[\rho_A^n] =& \int \prod_i d\alpha_id\bar \alpha_i d\beta_i d\bar\beta_i  
\nonumber \\
& \  \ \ \ \times\prod_i \left[ \rho_A(\bar\alpha_i,\beta_i) e^{-\sum_{i=1}^n (\bar\alpha_i\alpha_i +\bar\beta_i\beta_i)} \right]\nonumber \\
&  \  \ \ \ \times  \braket{\bar\beta_1|\alpha_2} \dots \braket{\bar\beta_{n-1}|\alpha_n} \braket{\bar\beta_n|-\alpha_1} \nonumber  \\
=&  \int \prod_i d\alpha_i d\bar \alpha_i \ \prod_i \left[ \rho_A(\bar\alpha_i,\alpha_i)\right] e^{\sum_{i,j} \bar\alpha_{i}  T_{ij} \alpha_j },
\end{align}
and the resulting twist matrix ${T}$ becomes (\ref{eq:Tmat}).

\begin{figure}
\includegraphics[scale=.55]{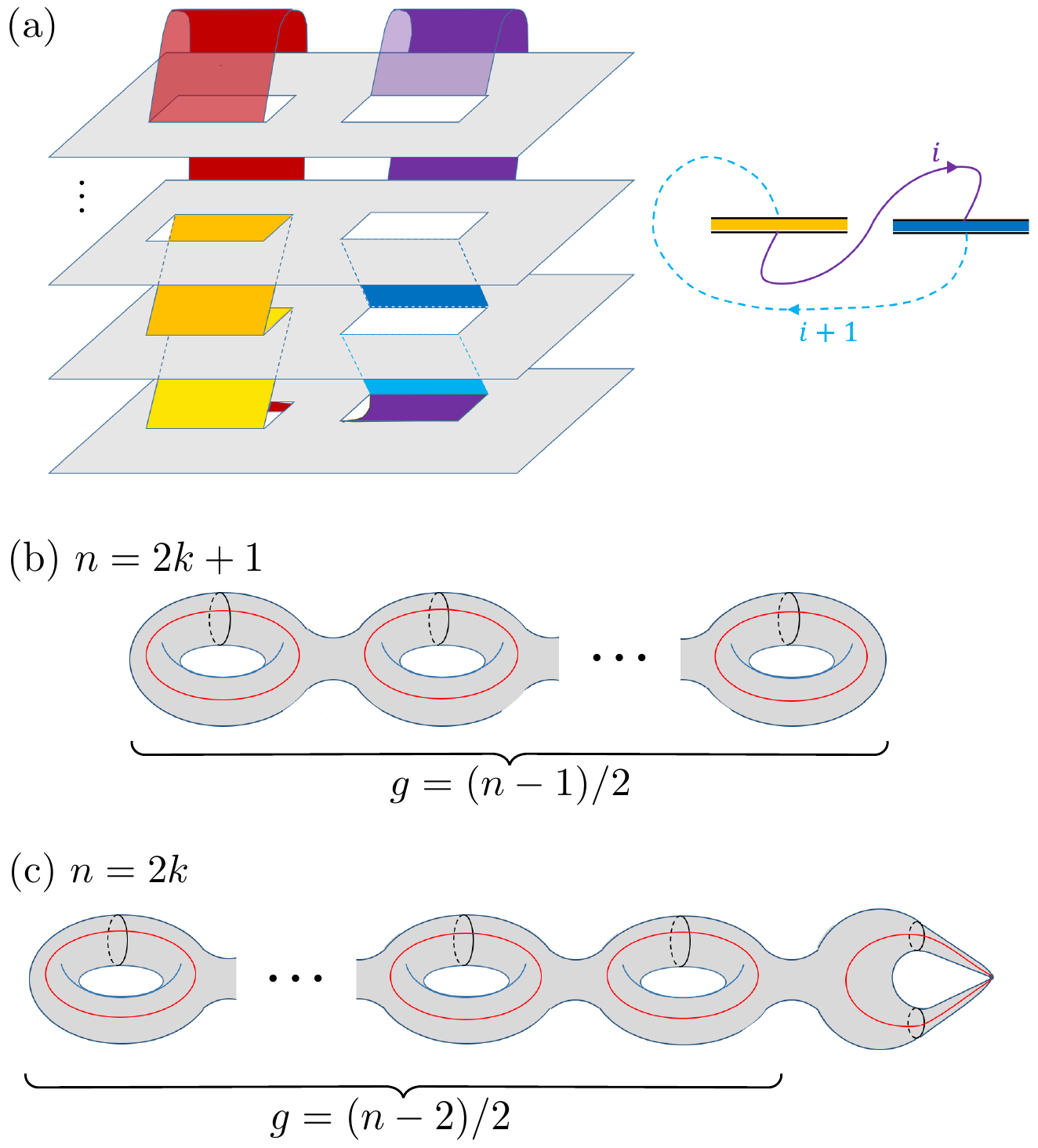}
\caption{\label{fig:OpOm} (a) Path integral representation of  moments of the partial time-reversal ${\cal E}_n\ $(\ref{eq:pathTR}). (b) and (c) Equivalent spacetime manifold of  ${\cal E}_n$ for two adjacent intervals  which is a torus $\Sigma_g$ of genus $g$ when $n$ is odd and $\Sigma_g$ with an additional pinched torus when $n$ is even. }
\end{figure}

We can also derive analogous relation for the TR operation. Let us recall the TR transformed density matrix
\begin{align}
\rho_A^{R}= \int d\alpha d\bar\alpha\ d\beta d\bar\beta\ \rho_A(\bar\alpha,\beta) \ket{i\bar\beta}\bra{i\alpha} e^{-\bar\alpha \alpha- \bar\beta\beta} ,
\end{align}
and
\begin{align} \label{eq:rhoTdag}
(\rho_A^{R})^\dag= \int d\alpha d\bar\alpha\ d\beta d\bar\beta\ \rho_A(\bar\alpha,\beta) \ket{-i\bar\beta}\bra{-i\alpha} e^{-\bar\alpha \alpha- \bar\beta\beta}.
\end{align}
Therefore, we can write for a product of $n$ density matrices (composed of $\rho_A^{R_1}$ and 
$\rho_A^{R_1\dag}$ alternating)
\begin{align}
&\Tr(\rho_A^{R}\rho_A^{R\dag}
\rho_A^{R}\rho_A^{R\dag}\cdots) \nonumber \\
=& \int \prod_i d\alpha_id\bar \alpha_i d\beta_i d\bar\beta_i  \nonumber \\
& \times \prod_i \left[ \rho_A(\bar\alpha_i,\beta_i) e^{-\sum_{i=1}^{n} (\bar\alpha_i\alpha_i +\bar\beta_i\beta_i)} \right] \nonumber \\
& \times \braket{i \alpha_1|-i\bar\beta_2} \dots \braket{i\alpha_{n-1}|-i\bar\beta_{n}} \braket{i\alpha_n|(-1)^{n}i\bar\beta_1}   \nonumber \\
=&  \int \prod_i d\alpha_i d\bar \alpha_i \ \prod_i \left[ \rho_A(\bar\alpha_i,\alpha_i)\right] e^{\sum_{i,j} \bar\alpha_{i} T^R_{ij} \alpha_j },
\end{align}
which means the TR operation is equivalent to the twist matrix
\begin{align} \label{eq:TRmat}
T^R=\left(\begin{array}{cccc}
0 & \cdots & 0 & (-1)^{n-1} \\
1 & \ddots & \vdots & \vdots  \\
0 & 1 & 0 & 0    \\
\cdots & 0 & 1 & 0 \\
\end{array} \right).
\end{align}
 Putting these together, we can write the general expression for moments of the partial TR as
\begin{align}
&\Tr(\rho_A^{R_1}\rho_A^{R_1\dag}
\rho_A^{R_1}\rho_A^{R_1\dag}\cdots) \nonumber\\
=& \int \prod_i d\alpha_i d\bar \alpha_i \ \prod_i \left[ \rho_A(\bar\alpha_i,\alpha_i)\right] 
 \nonumber \\
&\ \ \ \ \times e^{\sum_{i,j} \bar\alpha_{i}^{A_1} T^R_{ij} \alpha_j^{A_1} }
 e^{\sum_{i,j} \bar\alpha_{i}^{A_2}  T_{ij} \alpha_j^{A_2} } 
 \label{eq:pTR}
\end{align}
where $\alpha^{A_j}$ refers to the sites in the $A_j$ interval. Figure~\ref{fig:OpOm}(a) shows the spacetime picture of this quantity.

\begin{figure}
\includegraphics[scale=.52]{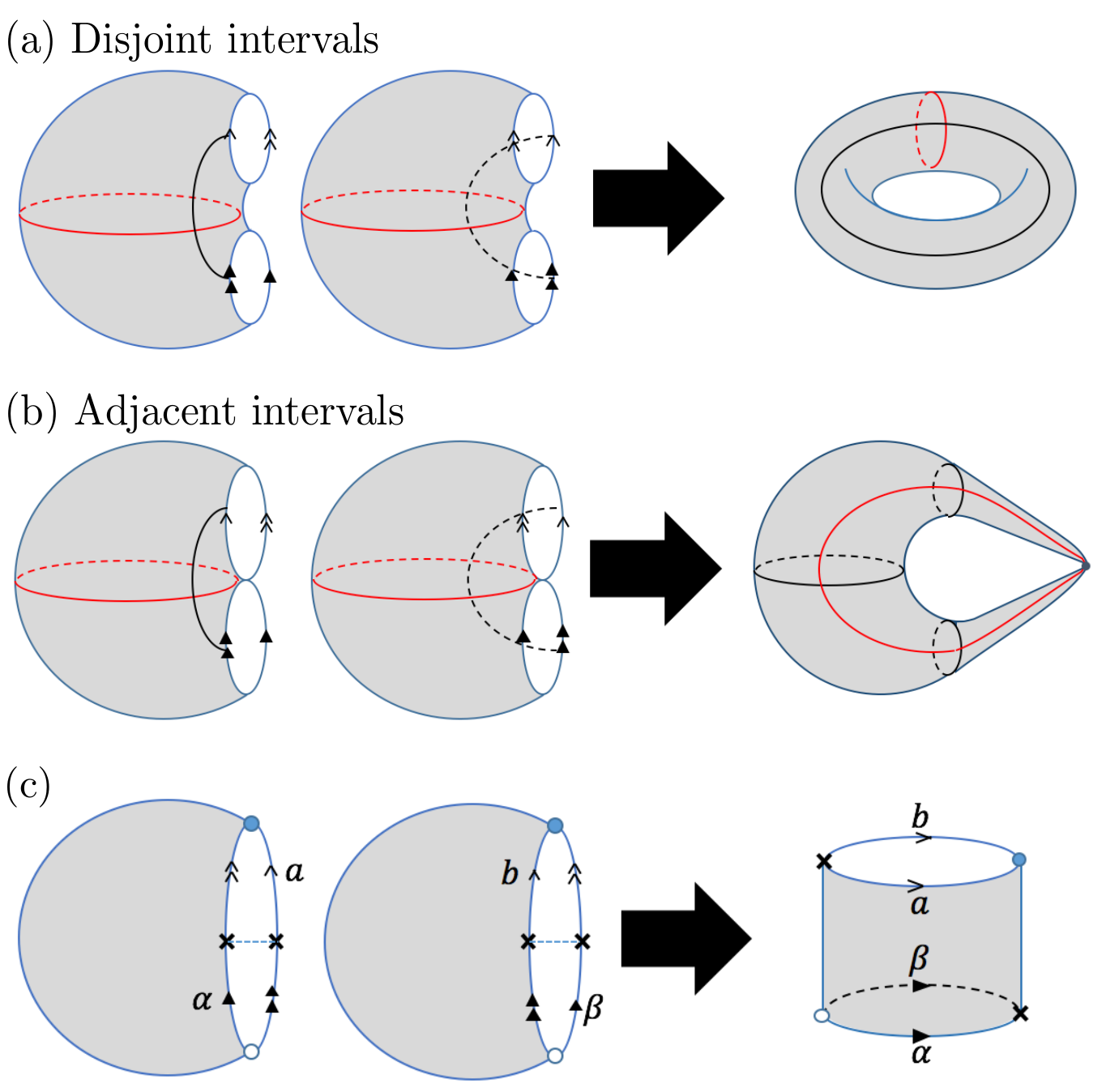}
\caption{\label{fig:rho2}  (Color online) Spacetime manifold of $\Tr [\rho_A^{R_1} \rho_A^{R_1\dag}]$ or  $\Tr [(\rho_A^{R_1})^2]$ for two disjoint intervals (a) and two adjacent intervals (b). In each panel, equivalent loops appear in the same color. (c) Spacetime manifold of $\rho_A^{R_1} \rho_A^{R_1\dag}$ or $(\rho_A^{R_1})^2$ for two adjacent intervals as a generator of ${\cal E}_n$ shown in Figs.~\ref{fig:OpOm}(b) and (c).}
\end{figure}

We now make a few remarks as to why the choice $[\rho^{R_1}_A\rho^{R_1\dag}_A\rho^{R_1}_A \cdots]$ is also a well-defined quantity from the quantum field theory point of view. First and foremost,  the above choice can be written as a partition function on the same spacetime manifold (Fig.~\ref{fig:OpOm}(a)) as the one originally proposed for the entanglement negativity in bosonic systems~\cite{Calabrese2013}. Moreover, only this quantity has a smooth behavior as we bring two disjoint intervals closer to each other to obtain two adjacent intervals. To illustrate this, let us consider the case of two replicas and the following quantities: $E_0=\Tr(\rho_A^2)$, $E_1=\Tr(\rho_A^{R_1}\rho_A^{R_1\dag})$, and $E_2=\Tr([\rho_A^{R_1}]^2)$. 
For two disjoint intervals, these quantities correspond to the toroidal spacetime (Fig.~\ref{fig:rho2}(a)), and for two adjacent intervals, $E_0$ defines the partition function on a sphere, while $E_1$ and $E_2$ define the partition function on a pinched torus (which is homeomorphic to a sphere with two points being identified, see Fig.~\ref{fig:rho2}(b)). 
Before we continue our discussion, it is worth mentioning that one can also consider another quantity $E_3=\Tr(\rho_A U_{A_1} \rho_A^{R_1} U_{A_1}^\dag)$. Although this quantity is a very useful measure in other contexts such as detecting the many-body topological invariants in time-reversal symmetric SPT phases~\cite{Shap2016,Shiozaki2016}, it is not qualified for our interest here, mainly for two reasons: First, it corresponds to the partition function on unorientable spacetime manifolds, while we want to consider orientable manifolds (Fig.~\ref{fig:OpOm}(a)) in analogy to what is considered for bosonic systems~\cite{Calabrese2013}.  Second, we are interested in an entanglement measure which can be defined for a generic system (possibly with no symmetry), while $E_3$ is only well-defined for time-reversal symmetric systems and vanishes otherwise. Returning to our earlier argument, we should note that the difference between $E_1$ and $E_2$ is in the boundary condition for the black loop in Fig.~\ref{fig:rho2}(b) (that is the loop between two successive replicas as shown in Fig.~\ref{fig:OpOm}(a)). $E_1$ and $E_2$ correspond to anti-periodic and periodic boundary conditions along such a loop, respectively.
As a result,  $E_2$ requires inserting a $\pi$-flux through the black loop around the pinched torus (to get a periodic boundary condition) whereas $E_1$ does not. 
However, this loop is contractible to a point as shown in Fig.~\ref{fig:rho2}(b), and 
this means that $E_2$ contains a singularity at the identification point. 
Hence, it is natural to choose $E_1$ (over $E_2$) that has no singularity in the limit of adjacent intervals.  
Given this observation, let us define moments of partial TR for any integer powers as follows
\begin{align} \label{eq:pathTR}
{\cal E}_n :=\left\{ \begin{array}{ll}
\ln \Tr (\rho_A^{R_1} \rho_A^{R_1\dag}  \cdots \rho_A^{R_1}\rho_A^{R_1\dag})& \ \ n\  \text{even}, \\
\ln \Tr (\rho_A^{R_1} \rho_A^{R_1\dag} \cdots \rho_A^{R_1}) &  \ \ n\  \text{odd}.
\end{array} \right.
\end{align}
The consecutive presence of $\rho_A^{R_1\dag}$ and $\rho_A^{R_1}$  implies a factor of $-1$ along the cycle between two replicas $i$ and $i+1$, as shown in Fig.~\ref{fig:OpOm}(a). The overall boundary condition along the twist operators is $T^n=(T^R)^n=(-1)^{n-1}$.  We find that the spacetime manifold for ${\cal E}_n$ is a higher genus torus shown in Figs.~\ref{fig:OpOm}(b) and (c) for even and odd values of $n$, where the boundary condition is anti-periodic for all the cycles. One way to see how these manifolds emerge is by noticing that the composite operator $\rho^{R_1}_A \rho^{R_1 \dag}_A$, for two adjacent intervals, forms a cylinder as shown in Fig.~\ref{fig:rho2}(c) and by combining these cylinders we construct the corresponding manifold for ${\cal E}_n$ which is sketched in Figs.~\ref{fig:OpOm}(b) and (c).

In the case of $n=2$, from the twist matrices $T$ and $T^R$ introduced above, it is easy to see that the identity 
\begin{align}
\Tr(\rho_A^2)=\Tr(\rho_A^{R_1}\rho_A^{R_1\dag})
\end{align}
holds in general.

\begin{figure}
\includegraphics[scale=.57]{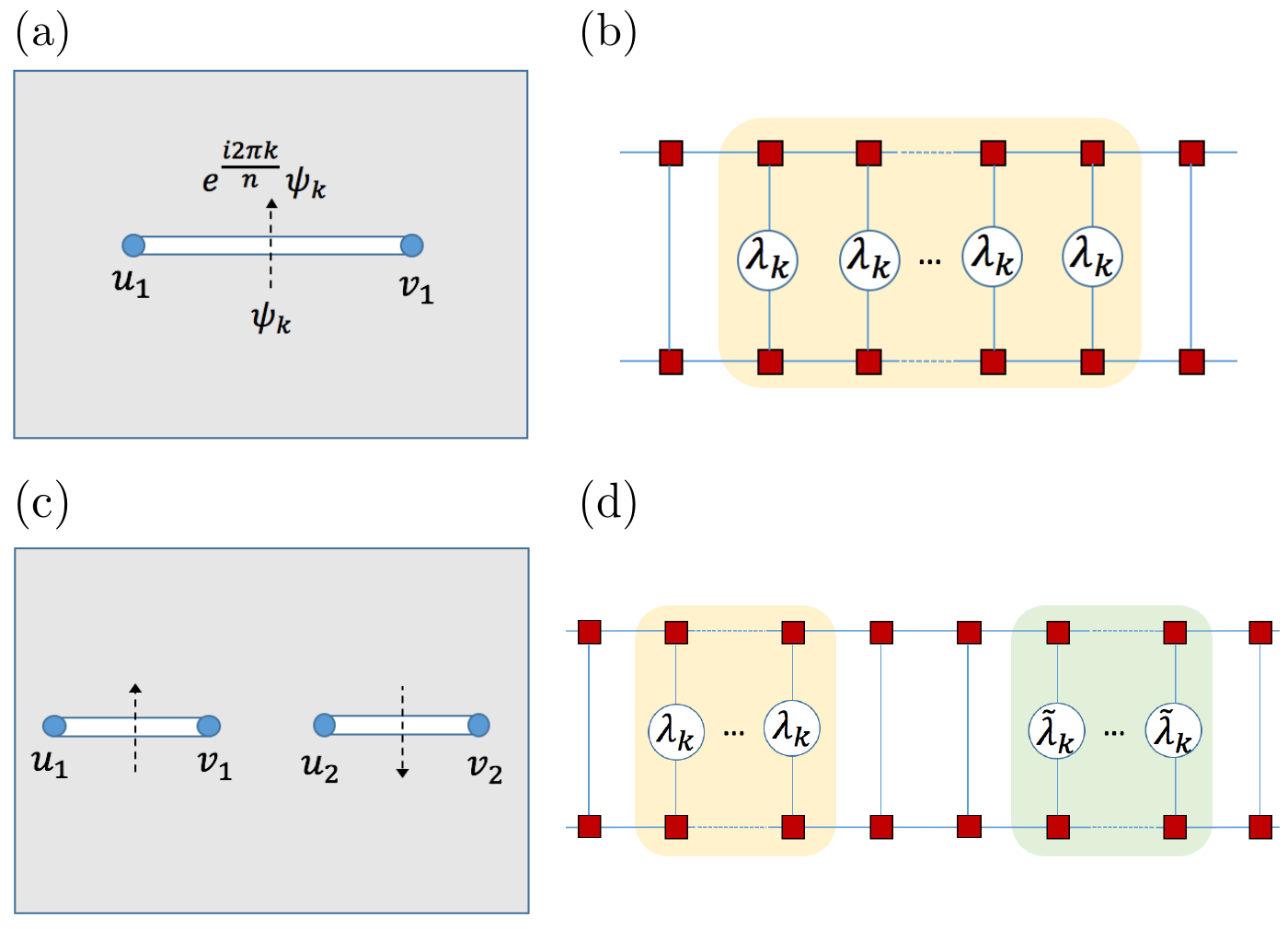}
\caption{\label{fig:negative} Path integral representation of R\'enyi entropy, (a) and (b), partial TR (\ref{eq:pathTR}), (c) and (d), in terms of $n$ decoupled partition functions with twist defects. (a) and (c) Spacetime picture, (b) and (d) MPS representation, $\lambda_k=e^{i2\pi k/n}$ and $\tilde\lambda_k=e^{i\delta} \lambda_k^\ast$ where $\delta=\pi$ or $\frac{\pi(n-1)}{n}$ for $n$ even or odd.}
\end{figure}

For generic noninteracting systems with conserved particle number, we can transform the trace formulas into a product of $n$ partition functions.
Let us first illustrate this idea for the REE~\cite{Casini2005}. We can diagonalize the twist matrix $T$ in Eq.~(\ref{eq:Tmat}) and rewrite the REE in terms of $n$-decoupled copies,
\begin{align}
\tr[\rho^n] =&  \int \prod_k d\alpha_k d\bar \alpha_k \ \prod_i \left[ \rho(\bar\alpha_k,\alpha_k)\right] e^{\sum_k \lambda_k \bar\alpha_{k} \alpha_k },
\end{align}
where $\lambda_k=e^{i 2\pi \frac{k}{n}}$ for $k=(n-1)/2,\cdots,(n-1)/2$ are eigenvalues of the twist matrix. In this new basis, the transformation rule (Eq.~(\ref{eq:Ttrans})) for the field passing through the interval becomes a pure phase, $\psi_k \mapsto \lambda_k \psi_k$  (Fig.~\ref{fig:negative}(a)). Therefore, the REE  can be decomposed into sum of separate terms as
\begin{align}
S_n = \frac{1}{1-n} \sum_{k=-(n-1)/2}^{(n-1)/2} \ln Z_k
\end{align}
where $Z_k$ is the partition function containing an interval with the twist phase $2\pi k/n$ (Fig.~\ref{fig:negative}(a)). 
In addition, $Z_k$ can be computed as a ground state expectation value,
\begin{align} \label{eq:Zk}
Z_k = \braket{\Psi|\hat T_k|\Psi}
\end{align}
where 
\begin{align} \label{eq:Tkop}
\hat T_k=\exp\left(i\frac{2\pi k}{n} \sum_{j\in A} f_j^\dag f_j
\right),
\end{align}
 is a phase twist operator which only acts on the $A$ interval.
This quantity can also be implemented in the  MPS representation as shown in Fig.~\ref{fig:negative}(b).

As shown in Eq.~(\ref{eq:pTR}), for the $n$-th moment of partially TR transformed density matrix 
${\cal E}_n$, we are dealing with two intervals where the twist matrices are $T$ and $T^R$. Fortunately, these matrices are simultaneously diagonalizable and we can reduce the $n$ coupled sheets to decoupled copies where the phase factors are different for the two intervals. Therefore, we can write
\begin{align}
 {\cal E}_n= \sum_{k=-(n-1)/2}^{(n-1)/2} \ln Z_{R, k}
\end{align}
where $Z_{R,k}$ is the partition function containing  two intervals with the twisting phases $e^{i2\pi k/n}$ and $e^{\delta-i2\pi k/n}$ (Fig.~\ref{fig:negative}(c)). This can also be implemented in the  MPS representation as shown in Fig.~\ref{fig:negative}(d), or equivalently,
\begin{align} \label{eq:ZRk}
Z_{R,k} = \braket{\Psi|\hat T_{R,k}|\Psi}
\end{align}
where 
\begin{align} \label{eq:TRkop}
\hat T_{R,k}=
\exp\left[
i\frac{2\pi k}{n} \sum_{j\in A_2} f_j^\dag f^{\ }_j 
+i(\delta-\frac{2\pi k}{n})\sum_{j\in A_1} f_j^\dag f^{\ }_j
\right], 
\end{align}
where $\delta=\pi$ or $\pi (n-1)/n$ for $n$ even or odd, respectively.
So far, we have presented three ways of computing the quantity ${\cal E}_n$: 
Eq.~(\ref{eq:app_f_21})
in terms of the many-body density matrix in the occupation number basis, 
Eq.~(\ref{eq:coh_moment}) in terms of coherent state basis, 
and the spacetime formula (\ref{eq:ZRk}) above. We numerically check that all these formulas are identical.


\section{\label{sec:examples} Examples}

In this section, we discuss the logarithmic negativity of two adjacent intervals in two canonical microscopic models: the Kitaev and Su-Schrieffer-Heeger chains. Our choice of models is motivated by the fact that these models offer three distinct regimes: Trivial phase where there is no entanglement between sites, Topological phase where nearby sites form singlet or Majorana bonds, and critical point which is described by CFT.
We would like to compare the resulting entanglement negativities due to the partial TR and the partial transpose in each regime.
In Appendix~\ref{sec:4x4}, we study a toy example of two-fermion density matrix which can also be used to represent the fixed-point density matrix in the topological and trivial limits of the above two models.

\subsection{the Kitaev chain}
As the first example, we apply our construction of the partial TR to the ground state of the Kitaev Majorana chain Hamiltonian~\cite{Kitaevchain}
\begin{align} \label{eq:BdG1d}
\hat{H}= 
-\sum_{j} \Big[t f_{j+1}^\dagger f_{j}+\Delta f_{j+1}^\dagger f^\dagger_{j} +\text{H.c.}\Big] -\mu \sum_{j} f_j^\dagger f^{\ }_j,
\end{align}
which describes a superconducting state of spinless fermions on a one-dimensional chain.
For simplicity, here we set $t=\Delta$. 
It should be noted that the topological phase with Majorana zero-energy edge modes 
is realized when $|\mu|/t < 2$.

\begin{figure}
\includegraphics[scale=1.1]{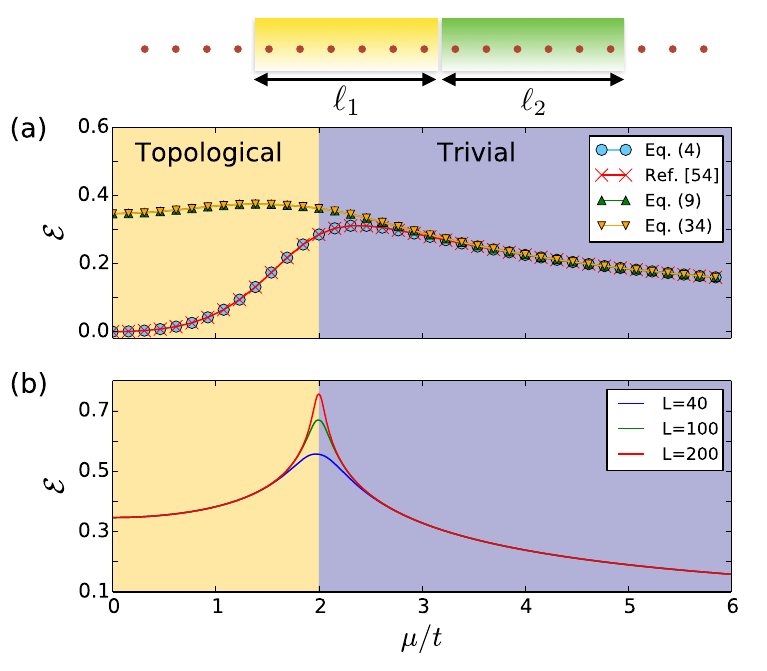}
\caption{\label{fig:Kitaev_vs_mu} (Color online) Logarithmic negativity of 
the Kitaev chain as a function of $\mu$ for two adjacent intervals with equal length $\ell$. (a) Comparison of different definitions of the partial transpose. From the legend, first curve (blue circles) is computed for the bosonic many-particle density matrix according to Eq.~(\ref{eq:app_f1}) in the Ising chain with periodic boundary condition.  Second (red crosses) and third (green upward triangles) curves are computed for the fermionic many-particle density matrix according to the rules introduced in Ref.~\cite{Eisler2015} and our paper Eq.~(\ref{eq:app_f_21}), respectively. 
Fourth curve (orange downward triangles) is computed using 
the free fermion formula,
Eq.~(\ref{eq:coh_neg}).
For all curves in this panel, we put $L=4\ell=8$. 
(b) Logarithmic negativity of the partial TR as computed in Eq.~(\ref{eq:coh_neg}) for large systems, $\ell=L/4$.
 All the data for fermionic chains are shown for  anti-periodic boundary condition.
  }
\end{figure}

Figure \ref{fig:Kitaev_vs_mu} shows the entanglement negativity of two adjacent intervals for various values of $\mu$.
As shown in Fig.~\ref{fig:Kitaev_vs_mu}(a), 
deep inside
the trivial phase both definitions of partial transformation consistently give zero.
On the other hand,
in the topological phase realized for $\mu/t<2$, 
and in particular near $\mu/t=0$, 
the partial transpose and partial time-reversal give very different results. 
For the fixed point wave function in the topological phase at $\mu=0$, 
the partial time-reversal yields $\mathcal{E}=\ln (\sqrt{2})$. 
This is expected since $A_1$ and $A_2$ share a Majorana bond,
which connects two Majoranas at the interface between the two intervals. 
On the other hand,
the other definition of the partial transpose in Ref.~\cite{Eisler2015} does not capture this and simply yields ${\cal E}=0$. 
In fact, we checked numerically that their result is equal to partial transpose of the bosonic density matrix (\ref{eq:parttrans_b}) in the equivalent Ising spin chain given by
\begin{align} \label{eq:Ising}
\hat{H}= 
-\sum_{j} [ J S^x_{j+1} S^x_{j} + h S^z_j],
\end{align}
where $S^x_j=(S^+_j+S^-_j)/2$ and $S^z_j= S^+_j S^-_j -1/2$ are spin-$1/2$ operators related to fermions through the Jordan-Wigner transformation
$S_j^-= \exp(i\pi \sum_{l<j} f^\dag_l f_l) f_l$
and
$S_j^+= \exp(-i\pi \sum_{l<j} f^\dag_l f_l) f_l^\dag$.
The Hamiltonians in Eqs.~(\ref{eq:BdG1d}) and (\ref{eq:Ising}) have identical ground state wave function, 
for values $t=\Delta=J/4$ and $\mu=h$, provided that we relate their bases so that spin-up state $\ket{\uparrow}=\sigma^+_j\ket{\downarrow}$ is identified with occupied state of fermion $f_j^\dag \ket{0}$ and spin-down $\ket{\downarrow}$ with an empty state $\ket{0}$.
Clearly, we should not expect any entanglement in the Ising chain in the limit $h\to 0$ 
as it describes a ferromagnetic ordered phase where there is no entanglement between neighboring sites. 
However, in the fermionic phase we have Majorana modes and the correct entanglement can only be captured 
in an intrinsic fermionic formalism as derived in this paper.
It is worth noting that in the Ising model language, the ground state for infinitesimally
small $h$ looks like an equal superposition of two ferromagnetically
ordered states, which gives rise to a contribution $\ln (2)$
to the (von Neumann) entanglement entropy for a finite interval
embedded in the whole system.
This correlation seen by the entanglement entropy
is however classical one,
and hence does not contribute 
to the entanglement negativity defined by using 
partial transpose. 


\begin{figure}
\includegraphics[scale=0.45]{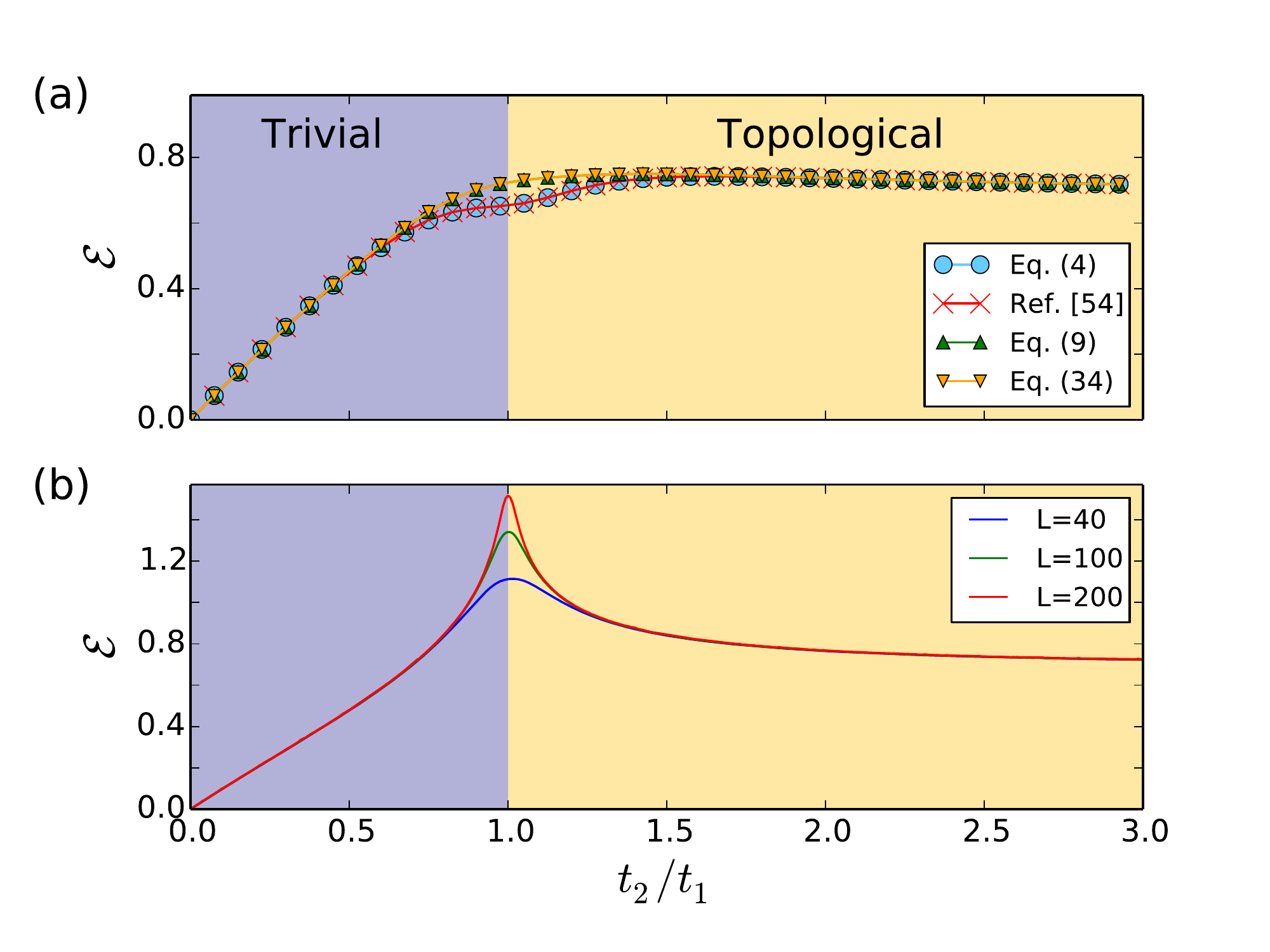}
\caption{\label{fig:SSH_vs_t2} (Color online) Logarithmic negativity of SSH model as a function of $t_2$ (\ref{eq:SSH}) for two adjacent intervals with equal length $\ell$. 
(a) Comparison of different definitions of the partial transpose.
See the caption of Fig.~\ref{fig:Kitaev_vs_mu} for details of what each curve represents. For all curves in this panel, we put $L=4\ell=8$ (16 fermion sites). 
(b) Logarithmic negativity of the partial TR as computed in Eq.~(\ref{eq:coh_neg}) for large systems, $\ell=L/4$. All the data for fermionic chains are shown for anti-periodic boundary condition. }
\end{figure}

\subsection{\label{sec:SSH} the Su-Schriffer-Heeger model}

As the second example, we consider the Su-Schrieffer-Heeger (SSH) model,
\begin{align} \label{eq:SSH}
\hat{H}= - \sum_j [t_2 f^{L\dag}_{j+1} f_j^R +t_1 f^{L\dag}_j f_j^R+ \text{H.c.}]
\end{align}
where there are two fermion species living on each site $f_j^L$ and $f_j^R$.
This model realizes two topologically distinct phases: topologically non-trivial phase for $t_2>t_1$, where the open chain has localized fermion modes at the boundaries, and trivial phase for $t_2<t_1$ which is just an insulator with no boundary mode. Figure~\ref{fig:SSH_vs_t2} compares the logarithmic negativity of two adjacent intervals using different methods. In the topological phase, both partial TR and fermionic partial transpose~\cite{Eisler2015} as well as bosonic partial transpose give $\ln(2)$ associated with the fermion bond at the sharing boundary between two intervals. 
The fact that the partial transpose can capture the entanglement in the SSH chain, but not the Kitaev chain, can be understood as a result of the violation of additivity (\ref{eq:additive}), see Appendix \ref{sec:Eisler} for explicit derivation. It is expected that any measure of entanglement $S$ which satisfies (\ref{eq:additive}), obeys the identity $S_{\text{SSH}}=2S_{\text{Kitaev}}$, since two copies of the Kitaev chain are equivalent to a single copy of the SSH chain. One way to see this is by fusing pairs of Majorana fermions across the two copies of the Kitaev chain to form complex fermions in the SSH chain.
Evidently, the negativity associated with the partial TR is consistent with this requirement.
Again, we observe that the fermionic partial transpose of Ref.~\cite{Eisler2015} is identical to the bosonic partial transpose. Here, the mapped bosonic Hamiltonian  after the Jordan-Wigner transformation is the $XY$ chain with alternating exchange coefficients,
\begin{align} \label{eq:XY}
\hat{H}= &
-\frac{t_2}{2} \sum_{j} [S^{L,x}_{j+1} S^{R,x}_{j} +S^{L,y}_{j+1} S^{R,y}_{j} ] \nonumber \\
& -\frac{t_1}{2} \sum_{j} [S^{L,x}_{j} S^{R,x}_{j} +S^{L,y}_{j} S^{R,y}_{j}],
\end{align}
and $\ln(2)$ in the bosonic partial transpose comes from breaking the spin bond $\ket{\uparrow\downarrow}+\ket{\downarrow\uparrow}$ at the interface between the two intervals.


\section{\label{sec:CFT} Critical points}
Calculation of the negativity at the critical point where the theory is 
conformally invariant has been a subject of great interest.
Here, we show that the entanglement negativity associated with the partial TR of two adjacent intervals obeys the familiar form 
\begin{align}
{\cal E}(\ell)=\ln \Tr |\rho_A^{R_1}|=\frac{c}{4}\ln \ell + \cdots
\end{align} 
where $c$ is the central charge of the conformal field theory (CFT)~\cite{Calabrese2013}. In the first part of this section, 
we analytically derive this result for massless Dirac fermions, which is the continuum theory for the critical point of the SSH model. To this end, we relate the negativity to the determinants of Toeplitz matrices and use
the Fisher-Hartwig conjecture to evaluate those determinants.
In the second part, we present numerical results which completely match the above expression.

\subsection{Toeplitz matrix and Fisher-Hartwig conjecture}

As we have shown in Sec.~\ref{sec:rep},
either ones of $\tr \rho_A^n$ or ${\cal E}_n$ (defined in (\ref{eq:pathTR})) can be written as a product of partition functions 
with partial twisted boundary conditions in time direction, which in turn can be recast in the form of ground state expectation value of phase twist operators (Eqs.~(\ref{eq:Tkop}) and (\ref{eq:TRkop})).
For massless Dirac fermions, the expectation values can be written in terms of Toeplitz matrices (see Appendix~\ref{sec:Toeplitz} for definition) and hence, their determinant can be computed using the Fisher-Hartwig conjecture~\cite{Jin2004}.
In what follows, we first illustrate these steps to calculate REE as a warm-up example and next, we discuss the negativity.

Let us consider the lattice realization of massless Dirac fermions 
(also known as the XX (or XY) chain),
\begin{align} \label{eq:XX}
\hat{H}= -t \sum_{j=0}^{L-2} f_{j+1}^\dag f_j +t f_0^\dag f_{L-1} + \text{H.c.}
\end{align}
at half filling with anti-periodic condition. 
Note that the ground state wave function is a Slater determinant,
\begin{align}
 \braket{\{r_j\}|\Psi}= \det (u_i (r_j))
 \end{align}
where $u_i (r_j)=\braket{r_j|u_i}$ refers to a set of single-particle eigenstates. Hence, each term in Eq.~(\ref{eq:Zk}) can be written as
\begin{align} \label{eq:Mdef}
Z_k = \det M^k_{mm'}= \det \left( \braket{u_m|T_k|u_{m'}}
\right)
\end{align}
in which $T_k$ is a single-particle matrix representation of $\hat T_k$ operator (\ref{eq:Tkop}),
\begin{align}
T_k = \text{diag}\left[1,1,\cdots,1,e^{i2\pi \frac{k}{n}}, \cdots,e^{i2\pi \frac{k}{n}}, 1, \cdots, 1
\right]
\end{align}
where unity entries correspond to the sites outside the interval and $e^{i2\pi \frac{k}{n}}$ entries correspond to the sites within the interval. 
The single-particle eigenstates of the Hamiltonian (\ref{eq:XX}) are just plane waves $u_m(r_j)=\frac{1}{\sqrt{L}} e^{i \frac{\pi m}{L} j}$ where 
$m=\pm,1,\pm 3,\cdots,\pm(\frac{L}{2}-1)$ and therefore, we get
\begin{align}
M^k_{mm'} &= \braket{u_m|T_k|u_{m'}}
\nonumber \\
&=\frac{1}{L} \sum_{j=0}^{L-1} e^{-i \frac{\pi j}{L} {(m-m')}} T_{k}(j) 
\nonumber \\ 
&=\frac{1}{2\pi} \int_0^{2\pi} d\theta\ e^{-i\theta {(m-m')/2}}\ T_k(\frac{L\theta}{2\pi}) 
\end{align}
where we drop second index of $T_k$ matrix and identify $T_{k}(j) \equiv T_{k,jj} $, since only diagonal elements are non-zero.  Notice that $(m-m')/2=0,1,2,\cdots, (\frac{L}{2}-1)$ which implies that the size of the matrix is $\frac{L}{2}\times\frac{L}{2}$. We assume both $L/2$ and $\ell$ are even.
We should note that the identity between the summation and integration is understood when $L\to \infty$, $\ell\to \infty$ while $ r=\ell/L$ is kept finite. 
 This shows that $M_{jj'}^k$ is a Toeplitz matrix (compare with Eqs.~(\ref{eq:Toeplitzdef1}) and (\ref{eq:Toeplitzdef2}) of Appendix~\ref{sec:Toeplitz}), where the generating function is given by
\begin{align}
\phi(\theta)=
 \left\{ \begin{array}{ll}
e^{i2\pi k/n} & -\pi r<\theta<\pi r, \\
1 &  \pi r<\theta<2\pi- \pi r. 
\end{array}\right. 
\end{align}
The generating function $\phi(\theta)$ has two
discontinuities at $\theta=\pm \pi r$ and it has the following canonical
factorization
\begin{align}
{\phi}(\theta)= \psi(\theta)
t_{\beta_1(k),\, -\pi r}(\theta)t_{\beta_2(k),\,
\pi  r}(\theta) 
\end{align}
 with
\begin{align}
\psi(\theta) &=e^{i2\pi  r k/n},
\nonumber \\
\beta(k) &= \beta_1(k)=-\beta_2(k)= -\frac{k}{n}. 
\end{align}
where the function $t_{\beta_r,\theta_r}(\theta)$ 
is defined in Eq.~(\ref{eq:tdef}). For $|k/n|<1/2$, we know that $|\text{Re} (\beta_1(k))
|<\frac{1}{2}$ and $| \text{Re} (\beta_2(k)) | < \frac{1}{2}$ and
the Fisher-Hartwig conjecture is a theorem~\cite{Basor1979}. Therefore, $Z_k= \det M_{ij}^k$ can be
asymptotically represented as
\begin{align}
Z_k=&\Bigl(2-2\cos(2\pi \ell/L)\Bigr)^{-k^2/n^2}  
\nonumber \\
&\times \left\{G\Bigl(1+\beta(k)\Bigr) G\Bigl(1-\beta(k)\Bigr)\right\}^2
e^{i\pi k \ell /n} \left(\frac{L}{2} \right)^{-2k^2/n^2}
\end{align}
where $G$ is the
Barnes $G$-function (Eq.~(\ref{eq:barnes}) of Appendix \ref{sec:Toeplitz}).
Hence, the only subsystem size-dependent terms are
\begin{align}
\ln \text{Tr} [\rho_A^n] 
=& -2 \left(\sum_{k=-(n-1)/2}^{(n-1)/2} \frac{k^2}{n^2}\right) \ln L \sin(\pi \ell/L)   \nonumber \\
=& - \left(\frac{n^2-1}{6n}\right)  \ln L\sin(\pi \ell/L) + \cdots
\end{align}
which leads to the well-known results
\begin{align}
S_n &= \frac{n+1}{6n} \ln \ell + \cdots
\end{align}
for the REE.
This result was also obtained by bosonizing the Dirac theory and also applied to the massive Dirac fermions~\cite{Casini2005}. In the bosonized theory, the intervals are realized by singularities (vortices) with winding number (vorticity) $\beta=2\pi k/n$ at their end points. However, we find that in the case of computing the entanglement negativity these vorticities do not give correct results when $|\beta_i|>1/2$ and must be modified, while there is no systematic way of modifying them in this construction.
As we discuss below, the Fisher-Hartwig method gives a unique way of modifying $\{\beta_i\}$ in these cases. It is interesting to note that if we put the modified $\{\widehat\beta_i\}$ (obtained from the Fisher-Hartwig method) in 
Casini's 
construction~\cite{Casini2005} we will then get the same (correct) results as the Fisher-Hartwig method.

For the negativity as in ${\cal E}_n$, we use Eq.~(\ref{eq:ZRk}) and write
\begin{align} \label{eq:MRdef}
Z_{R,k }= \det M^{R,k}_{ij}= \det \left( \braket{u_i|T_{R,k}|u_j}
\right)
\end{align}
in which $T_{R,k}$ matrix (\ref{eq:TRkop}) contains the two adjacent intervals,
\begin{align}
T_{R,k} = \text{diag} {\Big[}& 1,\cdots,1,e^{i\delta-i2\pi \frac{k}{n}}, \cdots, e^{i\delta-i2\pi \frac{k}{n}},e^{i2\pi \frac{k}{n}}, \cdots,
\nonumber \\
& e^{i2\pi \frac{k}{n}},1, \cdots, 1 {\Big]}.
\end{align}
Following what we did for the REE, here the generating function is found to be
\begin{align}
\phi(\theta)=
 \left\{ \begin{array}{ll}
e^{i\delta-i2\pi k/n} & -\pi  r_1<\theta<0 \\
e^{i2\pi k/n} & 0<\theta<\pi  r_2 \\
1 &  \pi  r_2<\theta<2\pi- \pi r_1 
\end{array}\right. 
\end{align}
where $ r_i=2\ell_i/L$.
There are three discontinuities in the generating function $\phi(\theta)$ in this case. Hence, it has the following canonical factorization
\begin{align}
{\phi}(\theta)= \psi(\theta) t_{\beta_1(k),\, -\pi r_1}(\theta)
t_{\beta_2(k),\,\pi  r_2}(\theta) t_{\beta_3(k),\,0}(\theta) 
\end{align}
 with
\begin{align*}
\psi(\theta) &=e^{i\pi r_1/2 - i\pi k (r_1-r_2)/n}, 
\\
\beta(k)&=\frac{k}{n}=
\beta_1(k)+\frac{\delta}{2\pi}=\beta_2(k), 
\nonumber \\
 \beta_3(k) &=-2\beta(k)+\frac{\delta}{2\pi}.
\end{align*}
We should note that $|\frac{k}{n}|<\frac{n-1}{2n}$ and the condition $| \text{Re} (\beta_3(k)) | < \frac{1}{2}$ is not fulfilled when $\frac{k}{n}<0$. This means that for $\frac{k}{n}<0$  the Fisher-Hartwig in its original form is not accurate. Fortunately, there is a trick~\cite{BasTr,Deift2011,Deift2013} to resolve this issue. The idea is to replace $\{ \beta_i \}$ by a new set of parameters $\{\widehat\beta_i= \beta_i + n_i \}$ where the integers $n_i$ (under the condition $\sum_i n_i=0$) are determined
such that the function 
\begin{align}
F_\beta= \sum_i (\beta_i +n_i)^2
\end{align}
is minimum. A simple inspection yields the following solutions for the minimum of $F_\beta$,
\begin{align}
\left\{
\begin{array}{ll}
(0,0,0) & \ \ \ \ \ \frac{k}{n} \geq 0 \\
(1,0,-1) & \ \ \ \ \ \frac{k}{n} < 0
\end{array}
\right.
\end{align}
where each tuple represents the values of $(n_1,n_2,n_3)$ in the corresponding range.

Thus, the Toeplitz determinant is given asymptotically by:
\begin{align}
Z_{R,k}&= 
\Bigl(2-2\cos(2\pi \ell_1/L)\Bigr)^{-(|k/n|-\delta/2\pi)(2|k/n|-\delta/2\pi)} \nonumber \\
&
\quad \times
\Bigl(2-2\cos(2\pi \ell_2/L)\Bigr)^{-|k/n|(2|k/n|-\delta/2\pi)} \nonumber \\
&
\quad \times
\Bigl(2-2\cos(2\pi (\ell_1+\ell_2)/L)\Bigr)^{|k/n| (|k/n|-\delta/2\pi)} \nonumber\\
&
\quad \times
J\left(k\right) \left(\frac{L}{2} \right)^{\Delta_k}, 
\end{align}
where $J \left( k\right)=
 \prod_{i=1}^{3} G(1+\beta_i(k))G(1-\beta_i(k)),$
in which $\beta_i$ must be replaced by $\widehat\beta_i$ when $k<0$,
and
\begin{align*}
\Delta_k =
-\frac{6k^2}{n^2}+\frac{3\delta}{2\pi} \left|\frac{k}{n}\right|-\frac{\delta^2}{2\pi^2} + 2(\frac{3k}{n}-\frac{\delta}{\pi}+1)\theta(-k)
\end{align*}
and $\theta(x)$ is the step function.
Therefore, the $\ell$-dependent terms are found to be in the form
\begin{align} \label{eq:fh_latt_neg}
{\cal E}_n
&=  c^{(1)}_n \ln L \sin(\pi\ell_1/L)+ c^{(2)}_n \ln L \sin(\pi\ell_2/L) 
\nonumber \\ 
&\quad 
+ c^{(3)}_n \ln L \sin(\pi(\ell_1+\ell_2)/L) + \cdots
\end{align}
where 
\begin{align}
c^{(1)}_{n_o} &=c^{(2)}_{n_o} =c^{(3)}_{n_o} = -\frac{1}{12} \left( n_o- \frac{1}{n_o} \right), 
\end{align}
for odd $n=n_o$, and
\begin{align}
c^{(1)}_{n_e} &=c^{(2)}_{n_e} = -\frac{1}{6} \left( \frac{{n_e}}{2}- \frac{2}{{n_e}} \right), \\
c^{(3)}_{n_e} &= -\frac{1}{6} \left( \frac{{n_e}}{2}+ \frac{1}{{n_e}} \right),
\end{align}
for  even $n=n_e$.
As a result, we can write 
\begin{align}
{\cal E}_n =\left\{ \begin{array}{l}
 - \left( \frac{n_o^2-1}{12n_o} \right) \ln \ell_1\ell_2 (\ell_1+\ell_2), \\
 - \left( \frac{n_e^2-4}{12 n_e} \right) \ln \ell_1\ell_2 - \left( \frac{n_e^2+2}{12n_e} \right) \ln (\ell_1+\ell_2)
\end{array} \right.
\end{align}
 in the continuum limit.
To find the coefficient of the negativity,  we analytically continue $n_e$ to $1$, which gives
\begin{align} \label{eq:cleanCFT}
{\cal E} 
&= \lim_{n_e\to 1} {\cal E}_{n_e}= \frac{c}{4} \ln \left( \frac{\ell_1\ell_2}{\ell_1+\ell_2} \right) + \cdots
\end{align}
and for the lattice model, we obtain
\begin{align}
{\cal E} 
&= \frac{c}{4} \ln \tan (\pi\ell/L) + \cdots
\label{eq:negCFT}
\end{align}
for $\ell_1=\ell_2=\ell$, where we restore the central charge $c$ in front of the logarithms.
As we can see, we recover all the same expressions as in Ref.~\cite{Calabrese2013} which were derived for the bosonic CFT.

\begin{figure}
\includegraphics[scale=0.44]{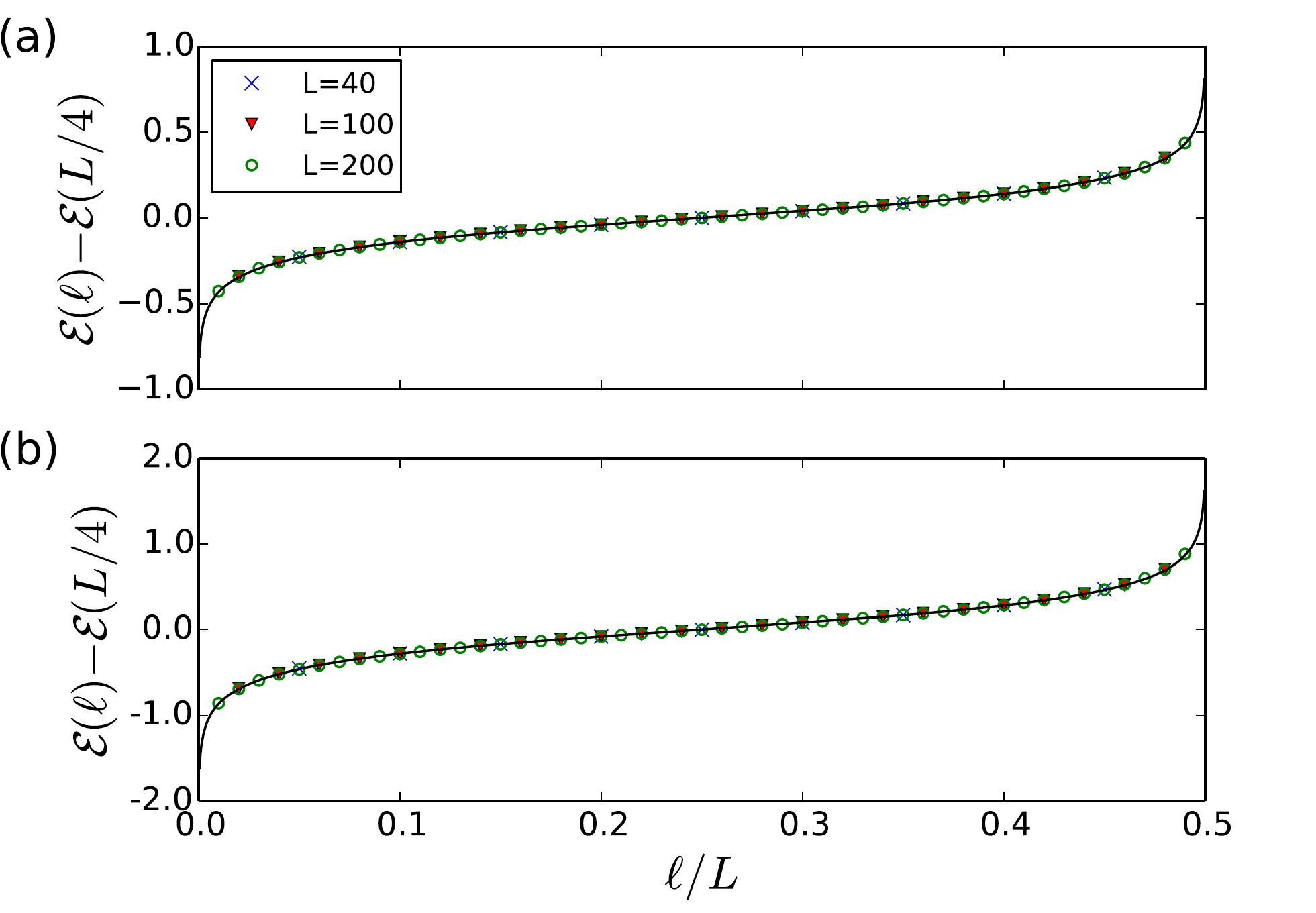}
\caption{\label{fig:Kitaev_CFT} (Color online) Logarithmic negativity of two adjacent intervals evaluated by Eq.~(\ref{eq:coh_neg}) at the criticality, (a) Kitaev chain (\ref{eq:BdG1d}) at $\mu=-2t$ where $c=1/2$, and (b) SSH model (\ref{eq:SSH})  at $t_2=t_1$ where $c=1$. 
The solid curves are the analytical results (\ref{eq:negCFT}). The data are shown for a closed chain with anti-periodic boundary condition.}
\end{figure}

\subsection{Numerical results}
In this part, we present numerical calculations for the critical systems in the clean limit and in the presence of random disorder. 
The numerical result for the clean limit conforms with the analytical result derived in the previous section. The result for the dirty limit also agrees with the previous studies of the negativity in random spin chains~\cite{Ruggiero_1}.

\subsubsection{Clean systems}
We check our analytical derivations in the previous part against numerical calculations in two cases:
the Kitaev chain
and SSH model at the criticality. The results are shown in Figs.~\ref{fig:Kitaev_CFT}(a) and (b), where the remarkable agreement is evident.

\begin{figure}
\includegraphics[scale=.35]{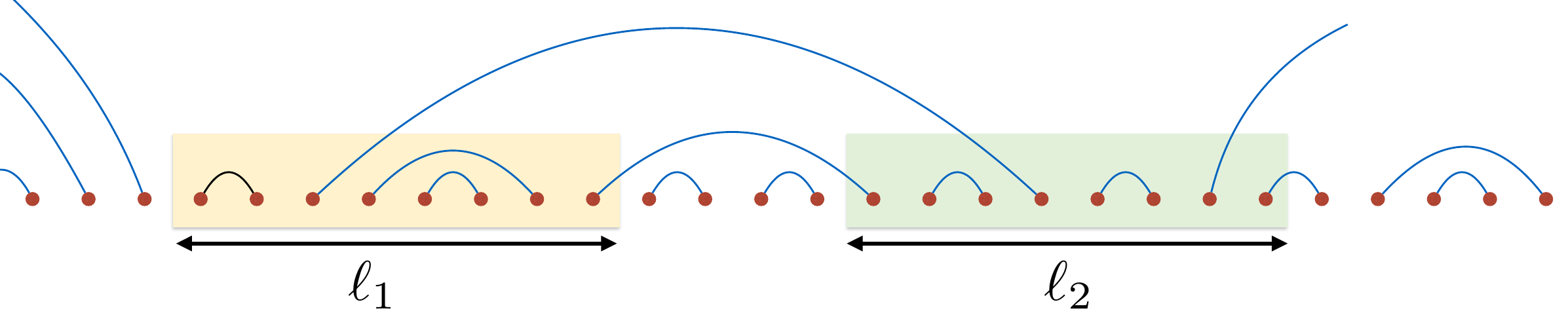}
\caption{\label{fig:RSP} The random singlet phase of disordered 
spin chains.}
\end{figure}

\subsubsection{Random singlet phase}

Now that we establish our construction of the negativity for fermions and provide a simple way of computing it for noninteracting systems, we would like to investigate the critical random spin chain.
We should note that strictly speaking,
we do not study the random spin chain, but the equivalent disordered fermionic chain with random hopping terms. Our goal is to give a quantitative evidence for the random singlet phase by using the entanglement negativity. In other words, we are interested in statistical distributions of singlet bonds by looking at the distribution of logarithmic negativity of two adjacent intervals.  
As we argue below, we expect that our fermionic calculation of the negativity yields similar results to the negativity of the random spin chain, which is found by the strong disorder renormalization group (SDRG)~\cite{Ruggiero_1}. First, we should note that 
the decimation scheme used in the SDRG to eliminate the high energy singlet bonds on a random spin chain can be similarly carried out in the fermionic chain (here to remove the nearest neighbor hopping) and the resulting RG flow equations for  the spin exchange coefficients and the hopping amplitudes are identical  \cite{Motrunich2002}. This means that on the fermionic side, SDRG gives rise to a random hopping model where the hopping amplitudes have identical distribution function to that of the spin exchange coefficients.
Second, the bosonic logarithmic negativity (in terms of partial transpose) returns $\ln (2)$ per each spin singlet $\ket{\uparrow\downarrow}-\ket{\downarrow\uparrow}$; analogously, the fermionic logarithmic negativity (in terms of partial TR) also gives $\ln (2)$ per each equivalent state $(f_1^\dag-f_2^\dag)\ket{0}$ (see Sec.~\ref{sec:SSH} and Appendix~\ref{sec:4x4}).
Hence, as far as the statistical distribution of logarithmic negativity is concerned, we do not anticipate any difference.
 In the following, we first briefly review the literature of the entanglement entropy in the critical random spin chains and then present our numerical results.


\begin{figure}
\includegraphics[scale=0.44]{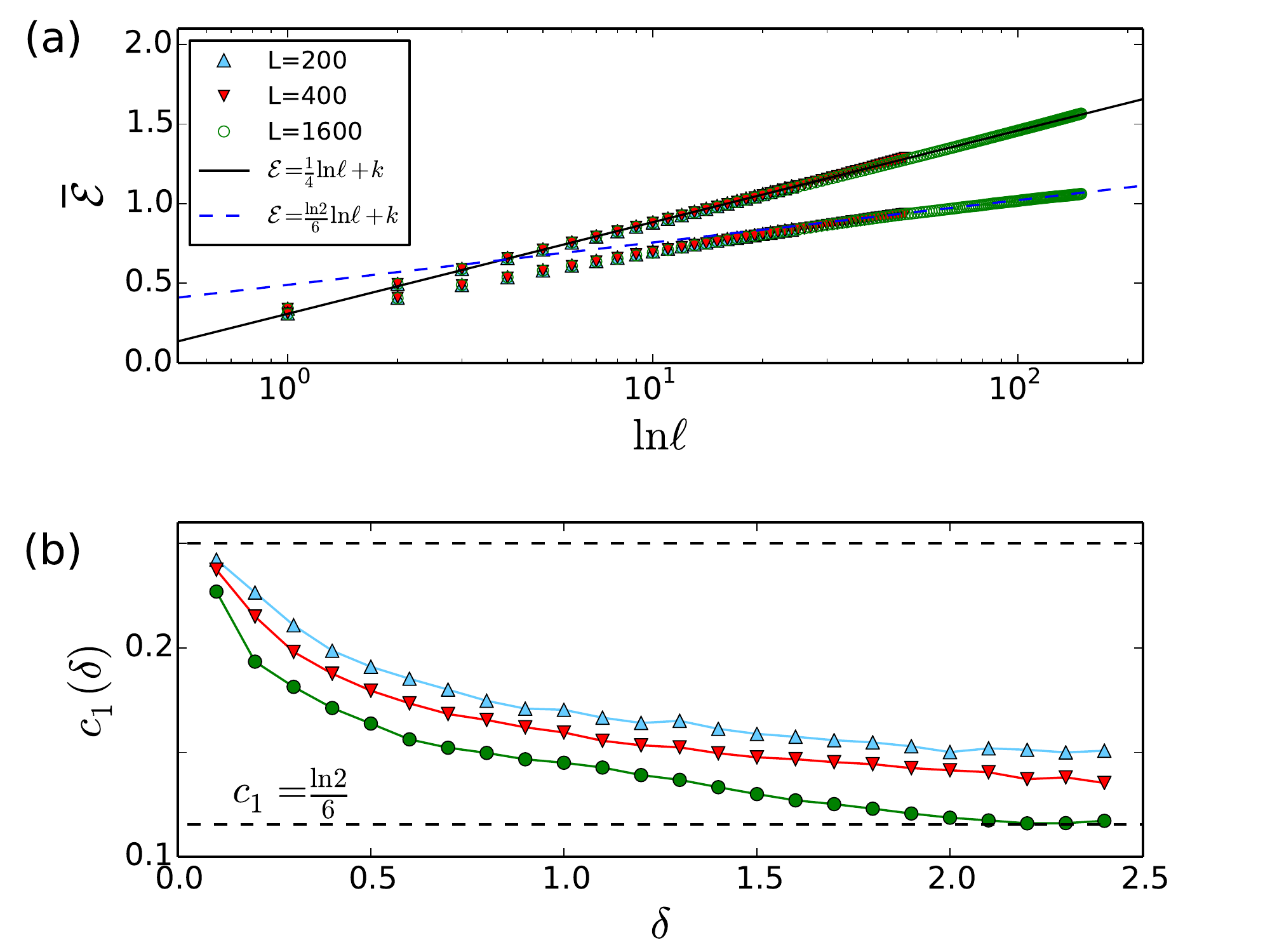}
\caption{\label{fig:rand} (Color online) 
(a) Logarithmic negativity of the random XX chain versus subsystem length, $\overline{\cal E}= c_1(\delta)\ln\ell +k $ for clean limit $\delta=0$ and strong disorder $\delta=2$. (b) The slope $c_1$ as a function of  disorder strength $\delta$ for various system sizes (see legend of (a)). The average is over $2\times 10^4$ realizations and error bars are smaller than symbol sizes.}
\end{figure}

We consider the random XX chain
\begin{align}
\hat{H}= &
\sum_{j} J_j [S^{x}_{j+1} S^{x}_{j} +S^{y}_{j+1} S^{y}_{j} ] ,
\end{align}
and its fermionic version
\begin{align} \label{eq:disXX}
\hat{H}= \frac{1}{2} \sum_{j} J_j (f_{j+1}^\dag f_j +f_{j}^\dag f_{j+1}).
\end{align}
The exchange coefficients (hopping amplitudes) $J_i$  are independent random variables taken from the distribution function
\begin{align} \label{eq:dist}
P_\delta (J) = \frac{1}{\delta} J^{-1+1/\delta}
\end{align}
defined over the range $J\in [0,1]$. The parameter $0\leq \delta<\infty$ is a measure of ``disorder strength'' and is introduced to tune between the clean limit where $\delta=0$, and the infinite-randomness fixed point (which will be defined shortly) where $\delta\to \infty$. Using the SDRG approach, it was shown that the low energy physics of this model 
(or generally, the XXZ chain) is described by the random-singlet phase (RSP) irrespective of the initial distribution function~\cite{Ma1979,Dasgupta1980,Fisher1994}. The fixed point limit of the RG flow leads to the distribution similar to $\delta\to \infty$ of $P_\delta(J)$ defined above and hence, it is named infinite-randomness fixed point (IRFP). The RSP is characterized by random singlet bonds between any two spins along the chain (see Fig.~\ref{fig:RSP}).
The signature of such random bonds can be captured by the von Neumann entanglement entropy~\cite{Refael2004} and R\'enyi entropies~\cite{Fagotti2011}. Interestingly, it was found that the disorder averaged von Neumann entropy of subsystem $A$ with length $\ell$ is proportional to the averaged number of singlet bonds between the subsystem and the rest of the system, $n_{A:B}$,
\begin{align}
\overline{S_{\text{vN}}} (\rho_A) &= -\overline{\Tr (\rho_A\ln\rho_A)} 
= \overline{n_{A:B}} S_{\text{vN}} (\rho_S) 
\end{align}
where the overbar refers to the average over disorder realizations, $\overline{n_{A:B}} \approx \frac{1}{3}\ln\ell$ and ${S_{\text{vN}}} (\rho_S) =\ln 2$ is the amount of entanglement per each singlet bond (denoted by the density matrix $\rho_S$).
Consequently, the entropy behaves as
\begin{align} \label{eq:disvN}
\overline{S_{\text{vN}}} (\rho_A)= \left(\frac{\ln 2}{3}\right) \ln \ell + \cdots
\end{align}
This result was then numerically confirmed for the XX chain (\ref{eq:disXX})~\cite{Laflorencie2005}.

The entanglement negativity of the RSP was recently studied~\cite{Ruggiero_1} and remarkably, a similar formula was derived
\begin{align} \label{eq:disform}
\overline{\cal E}= \overline{\ln \Tr |\rho_A^{T_1}|}= \overline{n_{A_1:A_2}} \ln 2
\end{align}
where this time the negativity is proportional to the average number of bonds between the subsystems $A_1$ and $A_2$. This expression further supports the intuition that the negativity is a measure of mutual entanglement.
 In particular, it was shown that the averaged negativity of two adjacent intervals  obeys the form
\begin{align} \label{eq:disCFT}
\overline{\cal E}= \left( \frac{\ln 2}{6} \right) \ln \left(\frac{\ell_1\ell_2}{\ell_1+\ell_2} \right) + \cdots .
\end{align}
Compared to the clean limit (\ref{eq:cleanCFT}), this is a striking result since the change in the coefficient is not merely a factor of $\ln 2$ as it was originally found for the von Neumann entropy (\ref{eq:disvN}).

We now present our numerical results for the negativity of two adjacent intervals and show that our numerics suggests this change of the coefficient from $1/4$ to $(\ln 2)/6$. It is worth remembering that Eq.~(\ref{eq:disCFT}) is only valid in the limit $\ell_i\ll L \to \infty$. In addition, the RSP fixed-point is a result of coarse-graining over a long chain and implies that $\ell_1$ and $\ell_2$ must also be sufficiently large. 
Numerically, we compute the negativity using Eq.~(\ref{eq:coh_neg}) and read off the slope of $\overline{\cal E}$ versus $\ln \ell$ where we consider equal lengths $\ell_1=\ell_2=\ell$ for the subsystems.
Instead of running computations over extremely large systems, our strategy is to simulate the transition by gradually tuning the disorder strength $\delta$ in the probability distribution (\ref{eq:dist}). In essence, tuning to larger values of $\delta$ is equivalent to moving deeper in the RG flow towards the RSP. In Fig.~\ref{fig:rand}(a), we show the averaged negativity for $\delta=2$ in comparison with the clean limit.
To gain more insight, we plot the slope of $\overline{\cal E}$  versus $\ln \ell$ as a function of disorder strength for various system sizes in Fig.~\ref{fig:rand}(b). There are two important observations here: The slope does change from $1/4$ to $(\ln 2)/6$ eventually at sufficiently large disorder and this transition gets sharper as we go to larger systems.

\begin{figure}
\includegraphics[scale=0.44]{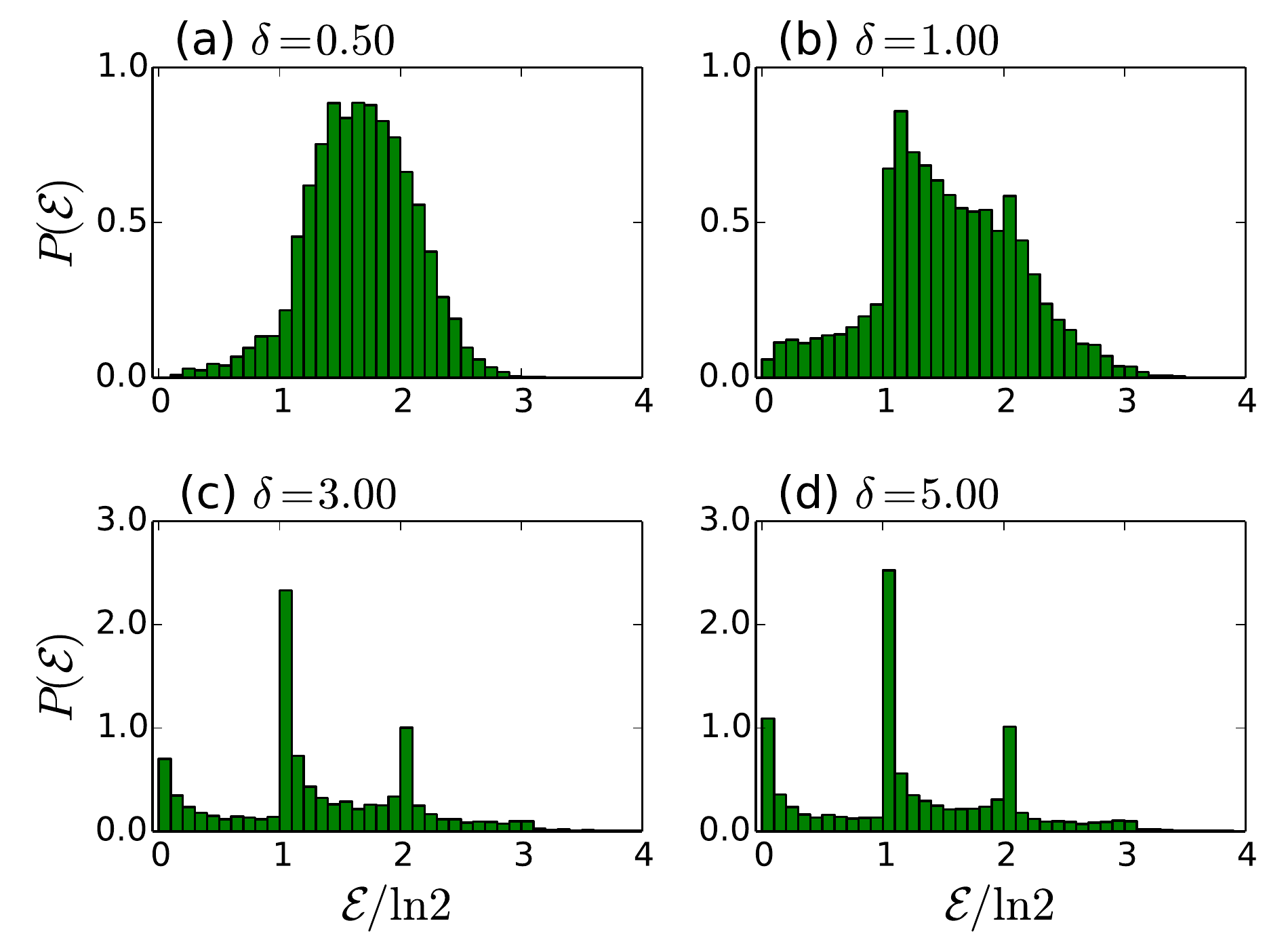}
\caption{\label{fig:hist} (Color online) Probability distribution $P({\cal E})$ of the entanglement negativity for 
the random XX chain with $L=8\ell=400$. We use $10^5$ samples to make these histograms.}
\end{figure}

A more direct signature of the RSP, compared to the slope of $\overline{\cal E}$ versus $\ln \ell$, can be illustrated by the distribution function of the entanglement negativity. In Fig.~\ref{fig:hist}, we make histograms of the negativity calculated for $10^5$ disorder realizations. The horizontal axis is normalized to $n_{A_1:A_2}={\cal E}/\ln 2$ to emphasize the functional form of the negativity as suggested by Eq.~(\ref{eq:disform}), which basically shows the distribution of 
the number of singlet bonds between $A_1$ and $A_2$. As it is evident in this figure, for small $\delta$ the distribution is rather broad and smooth; next, it starts to develop peaks at integers as we go to larger $\delta$. Finally, at extremely large $\delta$, the majority of the samples give integers, which is a strong evidence for the formation of only singlet bonds between the two subsystems.

\section{\label{sec:discussion} Discussion}
In conclusion, we propose a definition for the partial transpose in fermionic systems based on time-reversal transformation. We find that the partial time-reversal transformation implies a certain spacetime picture for the negativity of fermions. We explain how to construct partial time-reversal  for noninteracting fermions and how to evaluate the corresponding entanglement negativity. We also show that our formalism can capture the entanglement in the topological phase of the Kitaev Majorana chain which was inaccessible to the previous definition of the fermionic partial transpose.
We derive analytical expressions for the negativity of fermionic systems at the criticality and verify them numerically.  Remarkably, we recover the same results as the original CFT calculations~\cite{Calabrese2013}. 
We apply our framework to the random XX chain and present  some numerical evidence for the random-singlet phase from the perspective of the entanglement negativity.

In principle, the notion of fermionic partial transpose as a partial time-reversal transformation and the fact that the associated entanglement negativity can be straightforwardly computed in noninteracting fermions open many new avenues for research. In particular, some exciting future directions could be the study of finite-temperature systems, quench dynamics and entanglement growth, and the negativity in higher dimensional systems. In the case of two dimensional systems, it would be interesting to look at the result of partial time-reversal on various topologically ordered phases of matter such as the fractional quantum Hall effect.

Despite the fact that our treatment is quite general, throughout this paper we mostly focus on the adjacent intervals. One interesting direction is to consider disjoint intervals and compare the result of different definitions of the partial transpose. 

{\it Note Added.--} Recently, we became aware of a recent independent work~\cite{Eisert2016}, which provides tight upper and lower bounds to the entanglement negativity in fermionic systems. In their paper, they adopted the conventional definition of partial transpose for fermions~\cite{Eisler2015} and the partial time-reversal defined in our paper was considered as an upper bound in the case of noninteracting fermions. 

\acknowledgements
We acknowledge insightful discussions with Zolt\'an Zimbor\'as, Po-Yao Chang, Vincenzo Alba, Xueda Wen, and Hitesh Changlani.
We are also grateful to the participants of the workshop ``Entanglement in Quantum Systems''
held at the Simons Center for Geometry and Physics (Oct. 3-7, 2016).
Computational resources were provided by the Taub campus cluster at the University of Illinois at Urbana-Champaign.
This work was supported in part 
by the National Science Foundation Grant No.~DMR-1455296,
by the U.S. Department  of  Energy,  Office  of  Science,  
Office  of  Advanced Scientific Computing Research, 
Scientific Discovery  through  Advanced  Computing (SciDAC) program
under Award Number FG02-12ER46875.
KS is supported by JSPS Postdoctoral Fellowship for Research Abroad.



\appendix
\begin{widetext}
\section{\label{sec:Eisler} Prior definition of the fermionic partial transpose}

In this appendix, we review the results of Ref.~\cite{Eisler2015} and make connections between our findings and theirs. 
Using the density matrix introduced in Eq.~(\ref{eq:denmat}), the partial transpose for the subsystem $A_1$ is defined by~\cite{Eisler2015} 
\begin{align}
\mathcal{R}(
c^{\kappa_{m_1}}_{{m_1}} \cdots c^{\kappa_{2m_{\ell_1}}}_{2m_{\ell_1}})
= (-1)^{f(\underline{\kappa})} c^{\kappa_{m_1}}_{{m_1}} \cdots c^{\kappa_{2m_{\ell_1}}}_{2m_{\ell_1}}, \; \; 
\mathrm{where} \;\; 
f(\underline{\kappa})
= \left\{ \begin{array}{ll}
0 & \text{if}\ |\underline{\kappa}| \; \mathrm{mod} \; 4 \in \{0,1\},\\
1 & \text{if}\ |\underline{\kappa}| \; \mathrm{mod} \; 4 \in \{2,3\}.
\end{array} \right.
\label{eq:ferm_part_2}
\end{align}
Let $P_1$ be the parity operator 
 on subsystem $A_1$, and define the operators
$\rho_{+}=\case{1}{2}(\rho_A + P_1 \rho_A P_1)$ and 
$\rho_{-}=\case{1}{2}(\rho_A - P_1 \rho_A P_1)$. This clearly implies $\rho_A = \rho_{+} + \rho_{-}$.
Using the Majorana basis, 
$\rho_+$ and $\rho_-$ can be written as
\begin{align}
\rho_+&= \sum_{\substack{\underline{\kappa}, \underline{\tau} \cr |\underline{\kappa}| \; \mathrm{even}}} 
w_{\underline{\kappa}, \underline{\tau}} \, 
c^{\kappa_{m_1}}_{{m_1}} \cdots c^{\kappa_{2m_{\ell_1}}}_{2m_{\ell_1}} 
c^{\tau_{m'_1}}_{m'_1} \cdots c^{\tau_{2m'_{\ell_2}}}_{2m_{\ell_2}'} 
 \, , \\
\rho_-&= \sum_{\substack{\underline{\kappa}, \underline{\tau} \cr |\underline{\kappa}| \; \mathrm{odd}}} 
w_{\underline{\kappa}, \underline{\tau}} \, 
c^{\kappa_{m_1}}_{{m_1}} \cdots c^{\kappa_{2m_{\ell_1}}}_{2m_{\ell_1}} 
c^{\tau_{m'_1}}_{m'_1} \cdots c^{\tau_{2m'_{\ell_2}}}_{2m_{\ell_2}'} . \, 
\end{align}
By linearity of the partial transpose,  we have $\rho_A^{T_1}=\rho_+^{T_1}
+ \rho_-^{T_1}$ , 
and $\rho_\pm^{T_1}$ can be obtained using the definition~(\ref{eq:ferm_part_2}):
\begin{align}
\rho_+^{T_1} &= \sum_{\substack{\underline{\kappa}, \underline{\tau} \cr |\underline{\kappa}| \; \mathrm{even}}} 
(-1)^{|\underline{\kappa}|/2} w_{\underline{\kappa}, \underline{\tau}} \, 
c^{\kappa_{m_1}}_{{m_1}} \cdots c^{\kappa_{2m_{\ell_1}}}_{2m_{\ell_1}} 
c^{\tau_{m'_1}}_{m'_1} \cdots c^{\tau_{2m'_{\ell_2}}}_{2m_{\ell_2}'} \, , \nonumber \\
\rho_{-}^{T_1} &=\sum_{\substack{\underline{\kappa}, \underline{\tau} \cr |\underline{\kappa}| \; \mathrm{odd}}} 
(-1)^{(|\underline{\kappa}|-1)/2}w_{\underline{\kappa}, \underline{\tau}} \, 
c^{\kappa_{m_1}}_{{m_1}} \cdots c^{\kappa_{2m_{\ell_1}}}_{2m_{\ell_1}} 
c^{\tau_{m'_1}}_{m'_1} \cdots c^{\tau_{2m'_{\ell_2}}}_{2m_{\ell_2}'}\, . 
\label{pm_part_trans}
\end{align}
In the case of Gaussian states, the generalized Gaussian operators $O_+$ and $O_-$ are introduced, such that
$\rho^{T_1}_{+}= \frac{1}{2} (O_+ + O_-)$ and $\rho^{T_1}_{-}= \frac{i}{2} (O_- - O_+)$.
Thus, the partial transpose is decomposed into two terms
\begin{align}
\rho^{T_1}_A= \frac{1-i}{2} O_+ + \frac{1+i}{2} O_- .
\label{ptrho}
\end{align}

Let us now write the partial time-reversal in this notation. From Eq.~(\ref{eq:ptrans_Maj}), we have
\begin{align}
\rho_+^{R_1} &= \sum_{\substack{\underline{\kappa}, \underline{\tau} \cr |\underline{\kappa}| \; \mathrm{even}}} 
i^{|\underline{\kappa}|} w_{\underline{\kappa}, \underline{\tau}} \, 
c^{\kappa_{m_1}}_{{m_1}} \cdots c^{\kappa_{2m_{\ell_1}}}_{2m_{\ell_1}} 
c^{\tau_{m'_1}}_{m'_1} \cdots c^{\tau_{2m'_{\ell_2}}}_{2m_{\ell_2}'}\, , \nonumber \\
\rho_{-}^{R_1} &=i \sum_{\substack{\underline{\kappa}, \underline{\tau} \cr |\underline{\kappa}| \; \mathrm{odd}}} 
i^{(|\underline{\kappa}|-1)}w_{\underline{\kappa}, \underline{\tau}} \, c^{\kappa_{m_1}}_{{m_1}} \cdots c^{\kappa_{2m_{\ell_1}}}_{2m_{\ell_1}} 
c^{\tau_{m'_1}}_{m'_1} \cdots c^{\tau_{2m'_{\ell_2}}}_{2m_{\ell_2}'}\, . 
\end{align}
which are the same as (\ref{pm_part_trans}), up to the factor of $i$ appearing in front of $\rho_-^{R_1}$. Therefore, the partial time-reversal is given by
\begin{align}
\rho_A^{R_1} &= \frac{1}{2} (O_+ + O_-) + i \frac{i}{2} (O_- - O_+) =  O_+
\end{align}
which means the partial time-reversal can be put into a single Gaussian state. 
As we have seen in Eq.~(\ref{eq:rhoTdag}) of the main text, $\rho_A^{R_1\dag}$  is obtained by replacing $i$ with $-i$ in $\rho_A^{R_1}$,
\begin{align}
\rho_A^{R_1\dag} &= \frac{1}{2} (O_+ + O_-) - i \frac{i}{2} (O_- - O_+) =  O_-.
\end{align}
Therefore, the corresponding entanglement negativity is given by
\begin{align}
{\cal E}= \ln \Tr \sqrt{ \rho_A^{R_1} (\rho_A^{R_1})^\dag}= \ln \Tr \sqrt{O_+ O_-}.
\end{align}

In the rest of this appendix, we check whether the original definition (\ref{eq:ferm_part_2}) fulfills three natural expectations for the partial transpose at the operator level and show that it does not satisfy two of them. In contrast, partial time-reversal manifestly fulfills all of them.
The three natural conditions are as follows:
\begin{enumerate}[(i)]
\item
Subsequent partial transpositions for two complementary intervals are identified with the full transposition 
\begin{align}
(\rho^{T_1})^{T_2} = \rho^T.
\end{align} 
{\it Check:} This conditions is not satisfied. Consider for example,
\begin{align}
\rho= 1+ c_1 c_3 c_5 c_7
\end{align}
where $c_1$ and $c_2$ are in $A_1$ and $c_3$ to $c_8$ are in $A_2$. Hence, we have
\begin{align*}
\rho^{T_1}= 1+ \mathcal{R}(c_1) c_3 c_5 c_7 = 1+ c_1 c_3 c_5 c_7 = \rho,
\end{align*}
 and
\begin{align*}
(\rho^{T_1})^{T_2}= \rho^{T_2}= 1+ c_1 \mathcal{R}(c_3 c_5 c_7) = 1- c_1 c_3 c_5 c_7.
\end{align*}
 However,
\begin{align*}
\rho^{T}= 1+ \mathcal{R}(c_1 c_3 c_5 c_7) = 1+ c_1 c_3 c_5 c_7 ,
\end{align*}
 which means $(\rho^{T_1})^{T_2} \neq \rho^T$.

\item
 Applying the full transposition twice returns the original density matrix,
 \begin{align}
 (\rho^T)^T = \rho.
 \end{align}
{\it Check:} This condition is always satisfied.
  
 \item When considering $n$ flavors of fermions,
 \begin{align}
 (\rho_{1 } \otimes\cdots \otimes \rho_{n })^{T_1}= (\rho_{1 })^{T_1}  \otimes\cdots \otimes (\rho_{n })^{T_1}.
 \end{align}
{\it Check:} This condition is not fulfilled. For instance, consider two fermion flavors $A$ and $B$ and suppose the density matrices are
 \begin{align}
\rho_A= \frac{1}{4} (1- i c_{A1} c_{A4}),  \qquad \rho_B= \frac{1}{4} (1- i c_{B1} c_{B4}) .
\end{align}
Physically, this could be the reduced density matrix of two Kitaev Majorana chains ($A$ and $B$).
Further, assume that site $1$ is in $A_1$ subsystem and site $2$ is in $A_2$ subsystem.
Their partial transpose are given by
 \begin{align*}
\rho_A^{T_1}=\frac{1}{4} [ 1- i  \mathcal{R}(c_{A1}) c_{A4}]=\rho_A , \qquad \rho_B^{T_1}= \frac{1}{4} [1- i  \mathcal{R}(c_{B1}) c_{B4}]=\rho_B .
\end{align*}
The full density matrix is
\begin{align*}
\rho= \rho_A\otimes \rho_B =\frac{1}{16} ( 1 -  i c_{A1} c_{A4} -  i c_{B1} c_{B4} - c_{A1} c_{A4} c_{B1} c_{B4}),
\end{align*}
and its partial transpose is
\begin{align*}
\rho^{T_1} &=  \frac{1}{16} [1 -  i \mathcal{R}(c_{A1}) c_{A4} -  i \mathcal{R}(c_{B1}) c_{B4} +  \mathcal{R}(c_{A1}  c_{B1}) c_{A4} c_{B4}] \\
&=  \frac{1}{16} (1 -  i c_{A1} c_{A4} -  i c_{B1} c_{B4} + c_{A1} c_{A4} c_{B1} c_{B4}).
\end{align*}
Clearly, $\rho^{T_1}\neq \rho_A^{T_1} \otimes \rho_B^{T_1}$.
 We now compute the negativity using the partial transpose and show that it violates (\ref{eq:additive}): 
 For each chain, we have
\begin{align}
{\cal E}(\rho_A)= \ln \Tr |\rho_A^{T_1}|= 0, \qquad  {\cal E}(\rho_B)= \ln \Tr |\rho_B^{T_1}|= 0,
\end{align}
while for the combined chain we get
\begin{align}
{\cal E}(\rho_A\otimes\rho_B)= \ln \Tr |(\rho_A\otimes\rho_B)^{T_1}|= \ln (2).
\end{align}
Let us first show the single chain result by constructing the Hilbert space in terms of two complex fermions:
\begin{align}
f_{A1}^{\dag} = \frac{ c_{A1} +ic_{A2} }{2}, && 
f_{A2}^{\dag} = \frac{ c_{A3} + ic_{A4}}{2}.
\end{align}
The density matrix is then given by 
\begin{align}
\rho_{A} &=
 \frac{1}{4}[1-  i^2  (f_{A1}+f_{A1}^\dag)(f_{A2}-f_{A2}^\dag)]
 \nonumber\\
&= \frac{1}{4} \begin{pmatrix}
1 & 0 & 0 & 1 \\
0 & 1 & 1 & 0 \\
0 & 1 & 1 & 0 \\
1 & 0 & 0 & 1 \\
\end{pmatrix},
\end{align}
where the density matrix is represented in the basis $\{\ket{0}, f^{\dag}_{A1} \ket{0}, f^{\dag}_{A2} \ket{0}, f^{\dag}_{A2} f^{\dag}_{A1} \ket{0} \}$. In this basis, the partial transpose reads as
\begin{align}
\rho_{A}^{T_1}= & \frac{1}{4} \begin{pmatrix}
1 & 0 & 0 & 1 \\
0 & 1 & 1 & 0 \\
0 & 1 & 1 & 0 \\
1 & 0 & 0 & 1 \\
\end{pmatrix}, 
\end{align}
and thus, $\Tr|\rho_A^{T_1}|=1$.
In order to compute the full tensor product density matrix $\rho_A\otimes \rho_B$, we can either define the full $16\times 16$ Hilbert space or fuse two Majorana fermions in each interval (1 or 2) to construct the complex fermions. The first approach is a little bit cumbersome to present here, and we show the second method. Nonetheless, we checked the $16\times 16$ and the result is identical. Let us introduce the complex fermion operators 
\begin{align}
f_{1}^{\dag} = \frac{ c_{A1} +ic_{B1} }{2}, && 
f_{2}^{\dag} = \frac{ c_{B4} + ic_{A4}}{2}.
\end{align}
 The full density matrix then reads
 \begin{align}
 \rho &= \frac{1}{4}[ 1- i^2 (f_{1}+ f_{1}^\dag) (f_{2}- f_{2}^\dag)-
   i^2  (f_{1}- f_{1}^\dag) (f_{2}+ f_{2}^\dag)
  - i^2  (f_{1}+ f_{1}^\dag) (f_{2}- f_{2}^\dag)(f_{1}- f_{1}^\dag) (f_{2}+ f_{2}^\dag)]
    \nonumber \\
   &= \frac{1}{2} \begin{pmatrix}
1 & 0 & 0 & 1 \\
0 & 0 & 0 & 0 \\
0 & 0 & 0 & 0 \\
1 & 0 & 0 & 1 \\
\end{pmatrix}. 
 \end{align}
 It is worth noting that the above density matrix is identical to the reduced density matrix of the SSH fixed point which describes a singlet bond between two fermion sites.
Hence, we obtain the partial transpose,
\begin{align}
\rho^{T_1}
   &= \frac{1}{2} \begin{pmatrix}
1 & 0 & 0 & 0 \\
0 & 0 & 1 & 0 \\
0 & 1 & 0 & 0 \\
0 & 0 & 0 & 1 \\
\end{pmatrix},
\end{align}
 which gives $\Tr|\rho^{T_1}|=2$. Therefore, it should be clear now why the partial transpose works for singlet bonds but not Majorana bonds, since it violates the condition of additivity (\ref{eq:additive}).

\end{enumerate}

\section{\label{sec:OpOm} Calculation of the spectrum of $\rho_A^{R_1} \rho_A^{R_1\dag}$ for Gaussian states}

In this appendix, we explain how to find the eigenvalues of the  composite operator $\Xi=\rho_A^{R_1} \rho_A^{R_1\dag}$ and $\Xi^\alpha$ for $\alpha \in \mathbb{R}$.
We start from multiplying the two density matrix operators,
\begin{align} \label{eq:xi}
\Xi &=\frac{1}{{\cal Z}_\rho^2} \int {d} [\chi] d[\bar{\chi}] {d}[\xi] d[\bar{\xi}] \ e^{\frac{1}{2} \sum_{i,j\in A} \boldsymbol{\chi}_i^T {S}^{R_1}_{ij} \boldsymbol{\chi}_j } e^{\frac{1}{2} \sum_{i,j\in A} \boldsymbol{\xi}_i^T \bar{S}^{R_1}_{ij} \boldsymbol{\xi}_j }
\ket{\{ \chi_j \}_{j \in A} } \braket{ \{ \bar{\chi}_j \}_{j \in A}|\{ \xi_j \}_{j \in A} } \bra{ \{ \bar{\xi}_j \}_{j \in A}} 
\nonumber\\
&=\frac{1}{{\cal Z}_\rho^2} \int {d}[\chi]  {d}[\bar\xi]  \ e^{\frac{1}{2} \sum_{i,j\in A} (\chi_i,\bar{\xi}_i) \tilde{S}_{ij} (\chi_j,\bar{\xi}_j)^T } 
\ket{\{ \chi_j \}_{j \in A} } \bra{ \{ \bar{\xi}_j \}_{j \in A}} 
\end{align}
where $\bar{S}^{R_1}= U_S^{T\ast} S U_S^{\ast}$ as defined in (\ref{eq:Umat}) and
\begin{align}
e^{\frac{1}{2} \sum_{i,j\in A} (\chi_i,\bar{\xi}_i) \tilde{S}_{ij} (\chi_j,\bar{\xi}_j)^T }
=\int d[\bar{\chi}] {d}[\xi] e^{\frac{1}{2} \sum_{i,j\in A} (\boldsymbol{\chi}_i,\boldsymbol{\xi}_i)  {M}_{ij} (\boldsymbol{\chi}_j,\boldsymbol{\xi}_j)^T } 
\end{align}
is obtained after performing the Gaussian integral, with the kernel
\begin{align}
M=\frac{1}{2}
\left(\begin{array}{cc}
S^{R_1} & K \\
-K^T & \bar{S}^{R_1}
\end{array}
\right)
\end{align}
in the basis $(\chi_i,\bar\chi_i,\xi_i,\bar\xi_i)$, where the submatrix $K$, due to the inner product $\braket{ \{ \bar{\chi}_j \}_{j \in A}|\{ \xi_j \}_{j \in A} }$, is given by
\begin{align}
K=
\left(\begin{array}{cc}
0 & 0 \\
\mathbb{I} & 0 
\end{array}
\right).
\end{align}
Therefore, from Eq.~(\ref{eq:xi}) we can read off the correlators using the relation
\begin{align} 
[\tilde{\Gamma}]^{-1}&=
(\tilde S-i\sigma_2)^{-1}=
\left( \begin{array}{cc}
\tilde F^\dag & -\tilde C^T \\
\tilde C & \tilde F
\end{array} \right),
\end{align}
and find the corresponding Green function matrix $\tilde{G}$,
\begin{align} 
\tilde{G}&= 
\left( \begin{array}{cc}
1-\tilde{C}^T & \tilde{F}^\dag \\
\tilde{F} & \tilde{C}
\end{array} \right).
\end{align}
As a result, we can write $\Xi$ in the diagonal basis
\begin{align}
\frac{\Xi}{\Tr (\Xi)} &=  \prod_j \frac{e^{- \zeta_j g^\dag_j g_j}}{ (1+e^{-\zeta_j})} 
= \prod_j  (1-\lambda_j)\ e^{-\zeta_j g^\dag_j g_j},
\end{align}
where $g_j= \sum_i U_{ji} f_i$ are fermion operators in this basis and eigenvalues $\zeta_i$ are related to the eigenvalues of $\tilde G$ by $\zeta_i = \ln \left(\frac{1-\lambda_i}{\lambda_i} \right)$.
Hence, for the $\alpha$-th moment of $\Xi$ we get
\begin{align}
\ln \left(\frac{\Tr(\Xi^\alpha)}{\Tr(\Xi)^\alpha}  \right)
&=  \ln \left[\prod_j (1-\lambda_j)^\alpha \left(1+ \left(\frac{\lambda_j}{1-\lambda_j}\right)^\alpha \right)  \right]
=  \sum_j  \ln \left[ \lambda_j^\alpha + (1-\lambda_j)^\alpha   \right].
\end{align}



\section{\label{sec:4x4} Toy example of fermionic density matrix}

Here, we study the smallest nontrivial density matrix for two fermions. 
This density matrix can realize the trivial phase and topological phases of the Kitaev and SSH chains in the fixed-point (zero-correlation length limit).

Let us take a two-site system $f_1^\dag$ and $f_2^\dag$,
where the density matrix is $4\times4$ and can be represented in the basis $\{\ket{0}, f^{\dag}_1 \ket{0}, f^{\dag}_2 \ket{0}, f^{\dag}_1 f^{\dag}_2 \ket{0} \}$. A generic fermionic density matrix obeys the form
\begin{align}
\rho =
\begin{pmatrix}
\times &  0 & 0 &\times   \\
0 & \times  & \times  & 0 \\
0 & \times  & \times  & 0 \\
\times  & 0 & 0 & \times  \\
\end{pmatrix} 
\end{align}
where non-zero entries are shown as $\times$. This form of density matrix is dictated by the fact that only even fermion parity entries are allowed for the reduced density matrix of fermion-parity symmetric systems~\cite{PhysRevA.76.022311,Baculs2009,Zimboras2014}.
As an example, let us consider the density matrix
\begin{align}
\rho = p \ket{\psi_1}\bra{\psi_1} + (1-p)  \ket{\psi_2}\bra{\psi_2} 
\end{align}
where $\ket{\psi_1}=  f_1^\dag f_2^\dag \ket{0} +  a\ket{0} $ and $\ket{\psi_2}= (a f_1^\dag + f_2^\dag ) \ket{0}$.
This density matrix can be recast in the form 
\begin{align}
\rho = \frac{1}{\cal Z}
\begin{pmatrix}
p a^2 &  0 & 0 & p a  \\
0 & (1-p)a^2  & (1-p)a  & 0 \\
0 & (1-p)a  & (1-p)  & 0 \\
pa  & 0 & 0 & p  \\
\end{pmatrix},
\end{align}
where ${\cal Z}=1+a^2$ is the normalization constant.
The partial transpose is then given by
\begin{align}
\rho^{T_1} =\frac{1}{\cal Z}
\begin{pmatrix}
pa^2 &  0 & 0 & (1-p)a   \\
0 & (1-p)a^2 & pa  & 0 \\
0 & pa  & (1-p)  & 0 \\
(1-p)a  & 0 & 0 & p  \\
\end{pmatrix},
\end{align}
and the partial time-reversal is found by
\begin{align}
\rho^{R_1} =\frac{1}{\cal Z}
\begin{pmatrix}
p a^2 &  0 & 0 & i(1-p) a  \\
0 & (1-p)a^2  & ipa  & 0 \\
0 & ip a & (1-p)  & 0 \\
i(1-p)a  & 0 & 0 & p  \\
\end{pmatrix}.
\end{align}

Let us now consider various limits of this density matrix.
When $a=0$, we have a product state and  $\rho^{T_1}=\rho^{R_1}=\rho$. This corresponds to a trivial phase. It immediately implies that $\tr |\rho^{T_1}|=\tr |\rho^{R_1}| =1$ and the negativity vanishes in both definitions.
In the second case, consider $a=1$ which corresponds to maximally entangled states of $\ket{\psi_1}$ and $\ket{\psi_2}$ defined above. The negativity can be found easily in each definition
\begin{align}
\Tr |\rho^{T_1}| &= 1 +  |1-2 p|,  \\
\Tr |\rho^{R_1}| &= 2 \sqrt{2p^2-2p+1} 
\end{align}
Here, we have two important situations:

\textbullet$\ $ SSH fixed point: When $p=0$, both definitions yield $2$, consistent with a singlet bond between two sites.

\textbullet$\ $ Majorana fixed point: When $p=1/2$ (see Appendix F of \cite{Shap2016} for a proof). In this case, $\rho^{T_1}=\rho$ and $\Tr |\rho^{T_1}|=1$ and the partial transpose does not capture the Majorana bonds.
Remarkably, the partial TR gives $\sqrt{2}$ as it should be.


\section{\label{sec:Toeplitz} Toeplitz matrix and Fisher-Hartwig conjecture}
Here is a summary of the definition of Toeplitz matrix and useful expressions~\cite{Jin2004}.
The Toeplitz matrix $T_{L}[\phi]$ obeys the form,
\begin{align} \label{eq:Toeplitzdef1}
T_{L}[\phi]= (\phi_{i-j}), \qquad i,j=0,\cdots,{L}-1
\end{align}
where
\begin{align} \label{eq:Toeplitzdef2}
\phi_k=\frac{1}{2\pi} \int_0^{2\pi} \phi (\theta) e^{-{i} k \theta} {d} \theta
\end{align}
is the $k$-th Fourier coefficient of the generating function
$\phi(\theta)$. The
asymptotic behavior of the determinant of Toeplitz matrices with
singular generating function is also known as
the Fisher-Hartwig (FH) conjecture.
Suppose we have the following decomposition of the singular generating
function 
\begin{align}
\phi(\theta)=\psi(\theta) \prod_{r=1}^R t_{\beta_r,\,
\theta_r}(\theta) u_{\alpha_r,\,\theta_r}(\theta)
\end{align}
where
\begin{align}
t_{\beta_r,\,\theta_r}(\theta)&=\exp [-i\beta_r (\pi-\theta+\theta_r)], \qquad \theta_r<\theta <2\pi+\theta_r\label{eq:tdef}\\
u_{\alpha_r,\,\theta_r}(\theta)&=\Bigl(2-2\cos
(\theta-\theta_r)\Bigr)^{\alpha_r},\quad\quad \text{Re}[
\alpha_r]>-\frac{1}{2}\label{udf}
\end{align}
and $\psi(\theta)$ is a smooth non-vanishing
function with zero winding number. In the limit $L\to \infty$, the
determinant of $T_{L}[\phi]$ is given by
\begin{align}
\det T_{L}[\phi]= \left({\cal F}[\psi]\right)^{L}
\left(\prod_{i=1}^R {L}^{\alpha_i^2-\beta_i^2}\right)
{\cal E}[\psi, \{\alpha_i\}, \{\beta_i\},\{\theta_i\}]. \label{fh}
\end{align}
Here, ${\cal F}[\psi]=\exp \left(\frac{1}{2\pi} \int_0^{2\pi}\ln
\psi(\theta) {d} \theta\right)$ and
\begin{align}
{\cal E}[\psi, \{\alpha_i\},\{\beta_i\},\{\theta_i\}]
&={\cal E}[\psi]
\prod_{i=1}^R G(1+\alpha_i+\beta_i) G(1+\alpha_i-\beta_i)/G(1+2\alpha_i) 
\nonumber\\
&\quad \times 
\prod_{i=1}^R \biggl(\psi_-\Bigl(\exp({i}
\theta_i)\Bigr)\biggr)^{-\alpha_i-\beta_i} \biggl(\psi_+
\Bigl(\exp(- {i} \theta_i)\Bigr)\biggr)^{-\alpha_i+\beta_i}
\nonumber\\
&\quad 
\times \prod_{1\leq i \neq j \leq R}
\biggl(1-\exp\Bigl({i}
(\theta_i-\theta_j)\Bigr)\biggr)^{-(\alpha_i+\beta_i)(\alpha_j-
\beta_j)},
\end{align}
assuming that there exists a Weiner-Hopf factorization for $\psi(\theta)$, i.e.
\begin{align}
\psi(\theta)= {\cal F}[\psi]\,
\psi_+\Bigl(\exp({i}\theta)\Bigr)\,
\psi_-\Bigl(\exp(-{i}\theta)\Bigr).
\end{align}
In addition, ${\cal E}[\psi]=\exp(\sum_{k=1}^{\infty} k s_k s_{-k})$, in which $s_k$ is
the $k$-th Fourier coefficient of $\ln \psi(\theta) $, and $G$ is the
Barnes $G$-function defined by
\begin{align} \label{eq:barnes}
G(1+z)=(2\pi)^{z/2} e^{-(z+1)z/2-\gamma_E z^2/2}
\prod_{n=1}^{\infty} \{ (1+z/n)^n e^{-z+z^2/(2n)}\},
\end{align}
where $\gamma_E$ is Euler constant. The FH conjecture has not been proven
for a general case. However, there are various special cases for
which the conjecture was proven. The simplest case which applies to the R\'enyi entropy is when $R=2$, and we have $\alpha_1=\alpha_2=0$ and $\beta_1=-\beta_2=\beta$. The FH conjecture (Eq.~(\ref{fh})) was proven for the case $|\beta|<1/2$~\cite{Basor1979}. 

In the case of negativity of two adjacent intervals, we are dealing with $R=3$, $\alpha_i=0$, $\beta=\beta_1+1/2=\beta_2$ and $\beta_3=-2\beta+1/2$. Here, we  have $|\beta_3|>1/2$ when $\beta <0$ and the FH conjecture in its original form breaks down.
The resolution for a generic set of $\{\beta_i\}$, satisfying $\sum_i \beta_i=0$, was initially conjectured~\cite{BasTr} and later proven~\cite{Deift2011,Deift2013}. The idea is to replace $\{ \beta_i \}$ by a new set of parameters $\{ \widehat\beta_i \}$, (so-called the orbit of $\beta$) which are defined by
\begin{align}
O_\beta= \{\widehat\beta: \widehat\beta_j=\beta_j+n_j, \sum_{j=1}^R n_j=0 \}.
\end{align}
We look for a set of integers $\{ n_j \}$  which minimize the function
\begin{align}
F_\beta =\min_{\widehat\beta\in O_\beta} \left( \sum_{j=1}^R \widehat\beta^2_j \right),
\end{align}
for a given $\{ \beta_i \}$ and denote the corresponding set by ${\cal M}_\beta=\{ \widehat\beta\in O_\beta: \sum_{j=1}^R \widehat\beta_j^2=F_\beta \}$.
Refs.~\cite{Deift2011,Deift2013} have shown that the Toeplitz determinant can be evaluated by sum of these solutions
\begin{align}
\det T_{L}[\phi]= \sum_{\widehat\beta\in {\cal M}_\beta} \left({\cal F}[(\prod_{r=1}^R e^{in_r\theta_r}) \psi]\right)^{{L}}
\left(\prod_{i=1}^R {L}^{\alpha_i^2-\widehat\beta_i^2}\right)
{\cal E}[(\prod_{r=1}^R e^{in_r\theta_r})\psi, \{\alpha_i\}, \{\widehat\beta_i\},\{\theta_i\}],
\end{align}
where we substitute $\beta$ and $\psi(\theta)$ with $\widehat\beta$ and $\left(\prod_{r=1}^R e^{in_r\theta_r}\right) \psi(\theta)$ in Eq.\ (\ref{fh}), respectively.

\end{widetext}

\bibliography{refs}

\end{document}